\documentclass{aa}
\usepackage[varg]{txfonts}
\usepackage{natbib}
\usepackage{lscape}
\usepackage{booktabs,caption}
\usepackage[]{threeparttable}
\bibpunct{(}{)}{;}{a}{}{,}
\usepackage{subfig}
\AddToHook{begindocument/before}{\usepackage{hyperref}}

\newcommand{\kms}{$\rm km\,s^{-1}$}

\newcommand{\nucl}{${\rm _{nucl}}$}
\newcommand{\tot}{${\rm _{tot}}$}
\newcommand{\fit}{${\rm _{fit}}$}
\newcommand{\dvn}{$\Delta v_{90}$}

\begin{document}

\title{The $\alpha$-element enrichment of gas in distant galaxies}

\author{ Anna~Velichko\inst{\ref{inst1},\ref{inst2}}
\and Annalisa~De Cia\inst{\ref{inst3},\ref{inst1}}
\and Christina~Konstantopoulou\inst{\ref{inst1}}
\and C\'edric~Ledoux\inst{\ref{inst4}}
\and Jens-Kristian Krogager\inst{\ref{inst5}}
\and Tanita Ramburuth-Hurt \inst{\ref{inst1}}
}

\institute{ Department of Astronomy, University of Geneva, Chemin Pegasi 51, 1290 Versoix, Switzerland \email{anna.velichko@unige.ch}\label{inst1} 
\and 
Institute of Astronomy, Kharkiv National University, Sumska 35, Kharkiv, 61022, Ukraine\label{inst2}
\and
European Southern Observatory, Karl-Schwarzschild Str. 2, 85748 Garching bei M\"unchen, Germany\label{inst3}
\and
European Southern Observatory, Alonso de C\'{o}rdova 3107, Casilla 19001, Vitacura, Santiago, Chile\label{inst4}
\and
Centre de Recherche Astrophysique de Lyon, Univ. Claude Bernard Lyon 1, 9 Av. Charles Andr\'e, 69230 St Genis Laval, France\label{inst5}
}

\date{Received date /
Accepted date }

\abstract{ The chemical evolution of distant galaxies cannot be assessed from observations of individual stars, in contrast to the case of nearby galaxies. On the other hand, the study of the interstellar medium (ISM) offers an alternative way to reveal important properties of the chemical evolution of distant galaxies. The chemical enrichment  of the ISM is produced by all the previous generations of stars and it is possible to precisely determine the metal abundances in the neutral ISM in galaxies. The chemical abundance patterns in the neutral ISM are determined by the gas metallicity, presence of dust (the depletion of metals into dust grains), and possible deviations due to specific nucleosynthesis, for example, $\alpha$-element enhancements.}
{ We aim to derive the metallicities, dust depletion, and $\alpha$-element enhancements in the neutral ISM of gas-rich mostly-metal-poor distant galaxies (Damped Lyman-$\alpha$ absorbers, DLAs). Furthermore, we aim to constrain the distribution of $\alpha$-element enhancements with metallicity in these galaxies.}
{We collected  a literature sample of column density measurements of O, Mg, Si, S, Ti, Cr, Fe, Ni, Zn, P, and Mn in the neutral ISM of DLAs at redshifts of $0.60 < z < 3.40$.  We used this sample to define a golden sample of DLAs with constrained observations of Ti and at least one other $\alpha$-element. 
By studying the abundance patterns, we determined the amount of dust depletion, solely based on the observed relative abundances of the $\alpha$-elements. We then used the abundances of Fe-peak elements to determine the overall metallicity of each system, after correcting for dust depletion. In addition, we studied the deviations from the basic (linear) abundance patterns. 
We divided our sample into two groups of galaxies based on the widths of their absorption lines ($\Delta v_{90}$ above or below 100 \kms), which may be considered as a proxy for their dynamical mass. We characterised the distribution of the $\alpha$-element enhancements  as a function of metallicity for the galaxy population as a whole, by fitting a piecewise function (plateau, decline, plateau) to the data.}
{We observed systematic deviations from the basic abundance patterns for O, Mg, Si, S, Ti, and Mn, which we interpreted as $\alpha$-element enhancements and a Mn underabundance. The distribution of the $\alpha$-element enhancements with metallicity is different in the high-$\Delta v_{90}$ and low-$\Delta v_{90}$ groups of galaxies.
We constrained the metallicity of the $\alpha$-element knee for the high-$\Delta v_{90}$ and low-$\Delta v_{90}$ groups of galaxies to be $-$1.02$\pm$0.15 dex and $-$1.84$\pm$0.11 dex, respectively. The average $\alpha$-element enhancement at the high-plateau is [$\alpha$/Fe]=0.38$\pm$0.07 dex. On the other hand, Mn shows an underabundance in all DLAs in the golden sample of $-$0.36$\pm$0.07 dex, on average.}
{We have constrained, for the first time, the distribution of the $\alpha$-element enhancement with metallicity in the neutral ISM in distant galaxies. Less massive galaxies show an $\alpha$-element knee at lower metallicities than more massive galaxies. This can be explained by a lower star formation rate in less massive galaxies. If this collective behaviour can be interpreted in the same way as it is for individual systems, this would suggest that more massive and metal-rich systems evolve to higher metallicities before the contribution of SN-Ia to [$\alpha$/Fe] levels out that of core-collapse SNe. This finding may plausibly be supported by different SFRs in galaxies of different masses. Overall, our results offer important clues to the study of chemical evolution in distant galaxies.}


\maketitle

\section{Introduction}
\label{sec:intro}

The outcome of the evolutionary history of galaxies is recorded in the interstellar abundances of chemical elements. Observations of the interstellar medium (ISM) in galaxies of various types, differing in terms of mass, size, metallicity, and their respective evolutionary stages, may provide a key to understanding the processes taking place in galaxies  \citep[][]{Maiolino2019}. 

In the Milky Way (MW) and nearby galaxies, one of the ways to trace their evolutionary process is from observations of the detailed abundances of chemical elements in individual stars \citep[][]{McWilliam1997, Tolstoy2009}. Although it is impossible to obtain information about individual stars in distant galaxies, the chemical properties of their interstellar medium (ISM) can be analysed in great detail. The chemical abundance patterns in the ISM is the result of enrichment produced by all the previous generations of stars. These patterns make up a snapshot of the evolutionary history of the galaxy. 

One of the techniques to determine element abundances in the gas of faint high-redshift galaxies is based on the detection of its imprint on the spectrum of a background sources which may be a quasi-stellar object (QSO) or gamma-ray burst (GRB) afterglow \citep[][]{Prochaska1999, Prochaska2007, Ledoux2002, Vreeswijk2004, DeCia2012}. 
Damped Lyman-$\alpha$ absorbers (DLA) are defined as systems with high column densities of neutral hydrogen (N(H~I)$\geq 2\times 10^{20}$ cm$^{-2}$) that appear in absorption in QSOs or GRBs spectra.

The use of DLAs to study galaxies is important for several reasons.
First, due to the fact that they are observed in absorption, there is no selection effect with respect to luminosity of the DLA-galaxies and presumably masses. Hence, it is possible to detect in this way both the faintest low-mass and bright high-mass galaxies.
In our study, it is important to cover a wide range of galaxies as much as possible in terms of masses.

Secondly, due to the fact that hydrogen in DLAs is mainly in a neutral state (HI), the HI atoms shield the gas against ionisation \citep[e.g.][]{Viegas1995}, and other elements are also mainly in their dominant states, while the other states are negligible. For example, it is known that under such conditions, oxygen and nitrogen, like hydrogen, are neutral (OI, NI), while elements such as C, Mg, Si, Fe, and S are predominantly singly ionised \citep[CII, MgII, etc., see, e.g.][]{Hernandez2020}. Therefore, there is no need to apply ionisation corrections, which, in turn, can introduce large uncertainties. This makes it possible to accurately determine the element abundances from the column densities.

Third, unlike emission spectra, in which only very strong spectral lines can be detected, absorption spectra provide very detailed chemical composition even for faint high-z galaxies since, in this case, it is possible to make measurements from very weak line profiles.

The $\alpha$-element enhancement, [$\alpha$/Fe], and its evolution with the metallicity [M/H] is a strong diagnostic of the chemical evolution of galaxies \citep[e.g. ][]{Tinsley1979, Hill1995, Chiappini1997, Tolstoy2009, Kobayashi2020, Matteucci2021}. More details about [$\alpha$/Fe] evolution with metallicity are discussed in Section \ref{sec:alpha-elements}. Until recently, the [$\alpha$/Fe] has mostly been observed in stars in the local group \citep[e. g. ][]{McWilliam1997, Tolstoy2009, Matteucci2021}.  \citet{Cullen2021} observed [$\alpha$/Fe] in nearby galaxies from a combination of stellar and gas metallicities. In the gas, the measurements of [$\alpha$/Fe] are additionally complicated by the presence of dust, which  dramatically alters the observed abundances. Large fractions of refractory metals are missing from the gas-phase and are instead incorporated into dust grains, a phenomenon called dust depletion  \citep[e.g. ][]{Field1974, Savage1996, Jenkins2009, DeCia2016}. When dust depletion is negligible, it has been possible in the past to observe $\alpha$-element enhancement in a few least dusty, typically metal-poor DLA systems \citep[][]{Dessauges-Zavadsky2002, Ledoux2002b, Dessauges-Zavadsky2006, Cooke2011, Becker2012, DeCia2016}. Recently, \cite{Ramburuth-Hurt2023} studied the depletion patterns of individual gas components within DLA systems and discovered $\alpha$-element enhancements for systems with a wide range of dust depletions. In this study, we analyse the abundance patterns of a sample of DLAs out to z$\sim$4, correct for dust depletion, and measure the $\alpha$-element enhancements in the gas in the DLA galaxies. Furthermore, we split our sample in a group of lower-mass galaxies and one of higher-mass galaxies to investigate the evolution of the [$\alpha$/Fe] distribution with metallicity for these groups.

We refer to metal relative abundances [X/Y] defined as:
\begin{equation}
  {\rm  [X/Y] =  log\,\frac{N(X)}{N(H)} - log\,\frac{N(X)_\odot}{N(H)_\odot} }
,\end{equation}
where N(X)$_\odot$, N(Y)$_\odot$ are column densities  from \citet{Asplund2021}, following the recommendations of \citet{Lodders2009}, as reported in Table 1 of \citet{Konstantopoulou2022}.

\section{Features of the nucleosynthesis of the elements under study}

\subsection{The $\alpha$-elements} 
\label{sec:alpha-elements}

The study of the distribution of  [$\alpha$/Fe] with metallicity, [M/H], is fundamental to understand the chemical evolution of galaxies \citep[e.g. ][]{Tinsley1979, Hill1995, Chiappini1997, Tolstoy2009, Kobayashi2020, Matteucci2021} and can provide important information on the star-formation history (SFH) in a galaxy. It is well known that $\alpha$-elements (i.e. those that can be created by the $\alpha$-process, e.g. O, Mg, Si, S, Ca, Ti) are mainly produced inside massive stars (${\rm M \geqslant 8M_\odot}$) with lifetimes lower than 30 Myr. They are subsequently released into the interstellar medium (ISM) via core-collapse Type II supernovae. As a result, the ejected matter is enriched in $\alpha$-elements relative to Fe, until after about 1 billion years most SNe Ia begin to explode and produce more iron \citep[][]{Tinsley1979}.

Schematically, in simplified form, this behaviour can be plotted in coordinates [$\alpha$/Fe] -- [M/H], as  shown in Fig. \ref{fig:alpha-met-gal}. During the first phase, when only core-collapse SNe (CCSNe) contribute to the abundance of elements, [$\alpha$/Fe] is about constant and forms a plateau. The level of the high-$\alpha$ plateau depends on the initial mass function \citep[IMF, ][]{McWilliam1997} because a larger number of massive stars produce more $\alpha$-elements. In addition, the IMF may become more "top-heavy" with increasing star formation \citep[][]{Koppen2007}.

Although $\alpha$-elements enrichment is dominated by CCSNe, SNe Ia also can partially contribute to the abundances depending on the element. For instance, O and Mg are produced almost exclusively by CCSNe, while Si, S, Ca and Ti in the given order have an increasing contribution from SNe Ia, so that about 1/3 of titanium contained in the Sun has been produced in SNe Ia \citep[][]{Andrews2017}. In addition, according to \citet{Woosley1995, Cayrel2004}, O is mainly produced by central helium burning with some contribution of neon burning; Mg is mostly product of hydrostatic carbon burning in a shell and explosive neon burning; Si and Ca are formed in both the oxygen burning and neon burning shells; Ti by explosive oxygen burning. As a result of these differences in the formation of $\alpha$-elements, the level of the high-$\alpha$ plateau can systematically differ by 0.15~dex, if oxygen is not taken into account \citep[][]{Cayrel2004, Petitjean2008}. 

After SNe Ia start to explode, the abundance of $\alpha$-elements relative to iron gradually decreases, and the transition point from one regime to another is commonly referred to as the '$\alpha$-element knee' \citep[][]{McWilliam1997}. The position of the knee is sensitive to the early star formation rate (SFR). Observations at different redshifts show that the SFR increases with galactic mass \citep[][]{Katsianis2016}. If the SFR is high, then more metals have time to form during the first phase, resulting in a knee located at a higher [M/H] \citep[e.g.][]{Tinsley1979, McWilliam1997}. The position of the $\alpha$-element knee has been measured in stars in local galaxies, and it seems to depend on the mass of the galaxy, i.e. the knee is at lower metallicities for lower stellar masses \citep[][]{deBoer2014}. Because [$\alpha$/Fe] are decreasing after the knee, the plateau is followed by a decay with a certain slope. According to \citet{Andrews2017}, the latter can be sensitive to the gas outflow rate.

Then, after a certain period of time, the third phase of the [$\alpha$/Fe] evolution begins, when the contribution in the chemical yields of the SNe Ia compensates those of the CCSNe. At this point, the [$\alpha$/Fe] reaches a plateau at a value close to zero. In the case of the MW, the transition point is at [M/H] = $-$0.2 \citep[][]{McWilliam1997}, and the low-$\alpha$ plateau level is at [$\alpha$/Fe] = 0.05. To distinguish between the two knees mentioned above, we call them high-$\alpha$ and low-$\alpha$ knee for transition points from the high-$\alpha$ plateau to the decay and from the decay to the low-$\alpha$ plateau, respectively.

Figure \ref{fig:alpha-met-gal} shows the curves [$\alpha$/Fe] -- [M/H] we obtained using the release DR17 of the Apache Point Observatory Galactic Evolution Experiment (APOGEE, \cite{Majewski2017}) data for satellite galaxies of the MW:  Large and Small Magellanic Clouds (LMC and SMC, respectively), Sagittarius Dwarf Galaxy (Sgr), Fornax (Fnx), and the now fully disrupted Gaia Sausage/Enceladus (GSE) system. We selected stars belonging the galaxies according to the lists from \cite{Hasselquist2021}. For comparison, the curve for the MW given by \cite{McWilliam1997} is also shown. For details on APOGEE DR17 data fitting, we refer to  Appendix \ref{app:dwarf_gal_fit}.

Of course, this piecewise model is simplified compared to the natural behaviour. In fact, the evolution of galaxies of different types, sizes, and masses have their own peculiarities. For example, dwarf galaxies often have several separate bursts of star formation in their evolutionary history, in contrast to the more or less stable star formation in large galaxies such as the MW \citep[][]{Tolstoy2009, Atek2022}. Such a discontinuous star formation manifests itself in the [$\alpha$/Fe]--[M/H] behaviour in the form of bump on a low-$\alpha$ plateau \citep[see  e.g. the modelled tracks in Fig. 4 from][]{Hasselquist2021}. We see such a bump (or a hint of it) in the behaviour of [$\alpha$/Fe] for the all considered dwarf galaxies (see dark gray curves in Fig. \ref{fig:alpha-galaxies}).
The abundances of $\alpha$-elements  follow each other fairly well, that is, they make up a fairly homogeneous group; however, there are small differences in their production, such as the location of synthesis and dependencies on stellar mass and metallicity. 

\subsection{Zinc}
\label{sec:zinc}

Zinc is a volatile element that is often used as a tracer of metallicity in the ISM of various objects, assuming that it is not captured in dust grains.  In reality, the depletion of Zn is non-negligible. Therefore, the evaluation of metallicity using Zn can be erroneous, especially in highly dusty environments.

Zinc is often considered to be an ideal tracer of iron-peak elements, because in the Solar neighborhood [Zn/Fe] is about 0.0 for [Fe/H] between $-$2 and 0 dex \citep[e.g. ][]{Nissen2004, Saito2009}. However, the nucleosynthetic nature of Zn is complex since it is neither an iron-peak nor an $\alpha$-element. At [Fe/H] less than $-$2~dex [Zn/Fe] increases towards lower metallicities up to $\approx$0.5 dex \citep[][]{Nissen2007}. For the red giants in the MW bulge, at [Fe/H] > 0~dex [Zn/Fe] shows a decrease down to $\approx-$0.5 dex \citep[][]{Barbuy2015}. In dwarf galaxies the behaviour [Zn/Fe] versus [Fe/H] is different from that observed in the MW due to the different star formation history. Instead of a plateau, dwarf galaxies show a constant decreasing [Zn/Fe] over the entire observed metallicity range  \citep[][]{Skuladottir2017, Hirai2018}. This difference between the behaviour of [Zn/Fe] in massive and dwarf galaxies can be explained by the complicated production of Zn.

According to \citep[][]{Bensby2003, Nissen2011, Mishenina2011, Barbuy2015, Duffau2017}, the relation [Zn/Fe] versus [M/H] tends to behave partially like an [$\alpha$/Fe] but with a smaller amplitude. \citet{Sitnova2022} showed from non-LTE calculations for the MW stars that Zn and Mg do not strictly follow each other. Nevertheless, [Zn/Fe] is decreasing with metallicity because of various contributions from core-collapse SNe and SNe Ia, and their different timescales (as for $\alpha$-elements).

\citet{DeCia2024} analysed element abundances in the neutral ISM of the Small and Large Magellanic Clouds and observed a potential tendency of a deviation of Zn with respect to Fe up to 0.2 dex at lower metallicities and down to $-$0.2 dex at higher metallicities. If confirmed, this might be in agreement with the behaviour derived from stellar measurements described above. However, this still needs to be verified with further investigations with more data.

\subsection{Phosphorus}
\label{sec:phosphorus}

Phosphorus is thought to be mainly produced in massive stars during their hydrostatic carbon and neon burning phases \citep[][]{Woosley1995}. SNe Ia produce less significant amount of P compared to massive stars \citep[][]{Leung2018}. Little or no P is produced in AGB stars \citep[][]{Karakas2016}. According to \citet{Maas2019, Maas2022}, the behaviour of P with metallicity in the MW is most similar to that of the $\alpha$-elements, especially Mg. Over the [Fe/H] range from $-$1.0 to 0.2 dex [P/Fe] decreases from $\sim$0.6 dex to $\sim-$0.2. This is in agreement with the assumption that P is likely mostly produced in CCSNe.

\subsection{Iron group elements Cr, Fe, and Ni}
\label{sec:iron-group}

Iron-group elements such as Cr, Fe, Ni are mostly produced by SNe Ia \citep[][]{Nomoto1997, Kobayashi2020, Kobayashi2020b}. From observations of the MW stars [Cr/Fe] increases with metallicity from $\sim-$0.5 to $\sim$0 in the range $\sim-$4.2<[Fe/H]<$\sim-$1.5 \citep[][]{Frebel2010, Xing2019, Cayrel2004}.
From the APOGEE data, [Cr/Fe] and [Ni/Fe] are around 0 within the range of [Fe/H] from $\sim-$1.5 to $\sim$0.5 dex \citep[][]{Lim2022}. \citet{Bensby2003} find for thin and thick disk stars in the Solar neighborhood a slight overabundance of Ni with respect to Fe at metallicities below 0 and a small increase of [Ni/Fe] up to 0.1 dex above [Fe/H] = 0.

In nearby dwarf galaxies, LMC, SMC, Fnx, Sgr, GSE and Sculptor, Ni follows the MW trend in the metal-poor regime while for [Fe/H] $\geq-$1.5 it becomes underabundant down to $-$0.2 dex \citep[e.g, ][]{Tolstoy2009, Van-der-Swaelmen2013, Lemasle2014, Hasselquist2021}. From the abundance measurements in the neutral ISM of LMC and LMC, Ni is slightly underabundant relative to Fe at lower metallicities and increases up to 0 or a bit above at higher metallicities, while [Cr/Fe] is about 0 dex in the entire range of [Fe/H] \citep[][]{DeCia2024}.

\subsection{Manganese}
\label{sec:manganese}

Manganese is an iron-group element. 
Despite extensive studies by many authors \citep[e.g. ][]{Seitenzahl2013, Eitner2020}, the production of Mn remains uncertain. [Mn/Fe] has a different evolution compared to both other iron-peak elements and $\alpha$-elements \citep[][]{Mishenina2015}. Compared to $\alpha$-elements, core-collapse SNe produce much less Mn. Hence, at low metallicities, Mn deficiency with respect to Fe (or Mn underabundance) is observed. When type Ia SNe start exploding and producing more Mn \citep[][]{Nomoto1997}, the underabundance gradually disappears with an increasing [Fe/H] \citep[e.g. ][ for the MW]{Mishenina2015, Prantzos2005, Mikolaitis2017}. 

Moreover, according to \citet{Bergemann2008}, in stellar atmospheres Mn abundances may be affected by deviations from the LTE in such a way that the measured values may be underestimated by up to 0.5--0.6 dex. However, \cite{Mishenina2015} did not confirm these deviations which is explained by the lack of high-quality atomic data and, hence, the inability to build an adequate model of Mn atoms. Nevertheless, this should not affect the values obtained from ISM.

\begin{figure}
\centering
\resizebox{\hsize}{!}
   {\includegraphics{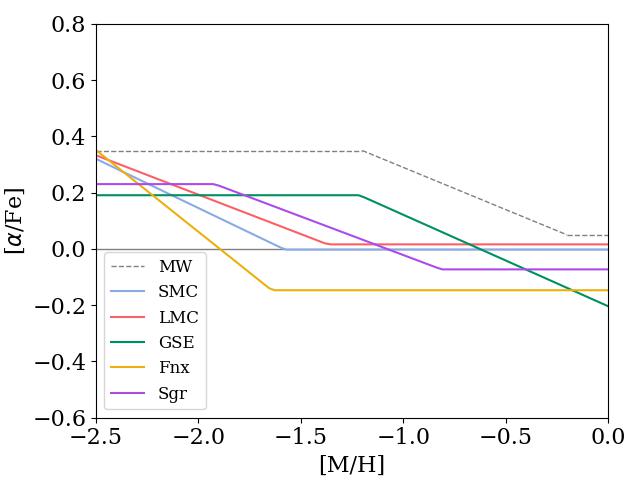}}
  \caption{Piecewise approximation of the $\alpha$-element behaviour in nearby dwarf galaxies SMC, LMC, GSE, Fnx, and Sgr from the APOGEE data (see details in \protect Fig. \ref{fig:alpha-galaxies}).}
\label{fig:alpha-met-gal}
\end{figure}

\section{Data and sample}
\label{sec:data}

\subsection{Column densities}
\label{sec:column_densities}

For the analysis, we utilised the column density measurements of O, Mg, Si, S, Ti, Cr, Fe, Ni, Zn, P, and Mn for the QSO-DLA sample taken from the work by \citet{Konstantopoulou2022}. After removing duplicate entries \citet[][]{Konstantopoulou2023} finalised a sample of 108 DLAs with a measurement of Zn. Among them, 37 QSO-DLAs are from \citet{DeCia2016}, which is a homogeneous sample of high-resolution ($R=\lambda/\Delta \lambda$ = 35000 -- 58000) spectra taken with the Very Large Telescope (VLT) Ultraviolet and Visual Echelle Spectrograph (UVES).  The signal-to-noise ratios (S/N) of these spectra are in the ranges 14--33 pixel$^{-1}$ and 54--10 pixel$^{-1}$ in the blue and red spectroscopic arm of UVES, respectively. The data were supplemented by \citet{Konstantopoulou2022}, wherever possible, with new column density measurements of Ti and Ni from UVES/VLT spectra using the Voigt-profile fit method \citep[][]{Krogager2018}. 71 QSO-DLAs are from the large compilation by \citet{Berg2015} published during the period between 1994 and 2004, with a minimum resolution restriction of $R >$ 10000, but typically $R \sim$ 40000. The S/N for the spectra is up to $\sim$50 pixel$^{-1}$, with the typical spectrum having an S/N of $\sim$10 pixel$^{-1}$ \citep[][]{Berg2015, Berg2015a}. In order to homogenise the measurements taken from different sources, \citet{Konstantopoulou2022} corrected the column densities to the newest possible  oscillator strengths for both of the samples.

The sample described above has been supplemented with one QSO-DLA (Q1210+175) from \citet{DeCia2016}, which is not included the work of \cite{Konstantopoulou2022} due to the lack of measurements for zinc; however UVES data covering absorption lines of Ti were indeed available. In addition, there is one QSO-DLA (Q1232+082) from the list provided by \cite{Konstantopoulou2022}, for which there was no Ti measurement, but it turned out to be possible. In this system, the Ti II $\lambda$1910 line can be slightly blended, but we consider the column density to be reliable. In this paper we measure Ti and Ni for these additional two systems, using the method described in \citet{Konstantopoulou2022}. In additional, we provide column densities of Ti and Ni, where is possible, for other DLAs from the list from \citet{DeCia2016} (see Table \ref{tab:col_dens}). Velocity profiles are shown in Figs. \ref{fig:col_dens_Q1210}, \ref{fig:col_dens_Q1232}. In total, our sample contains 110 QSO-DLAs.

From the list of 110 QSO-DLAs described above, we have selected 24 objects for which there are column density measurements of Ti and at least one more $\alpha$-element (Mg, Si, S, and O). We did not take into account those objects for which only limits have been calculated. We refer to this subsample as the "golden" sample. The need for this selection is explained in Section \ref{sec:method}. 15 QSO-DLAs in the golden sample are from the list of \citet{DeCia2016}, while the remaining 9 QSO-DLAs are from the compilation by \citet{Berg2015} (see references in Table \ref{tab:met_golden}). All the data for the golden sample is high quality, with resolution of at least 35000 and high S/N. 

The results of this work are based on the golden sample. For the rest of the sample, we do a minimal analysis and report the abundance patterns in the appendix.

\subsection{Velocity widths}
\label{sec:data_dv}

In addition to the column densities, our analysis uses information about the velocity widths of the absorption lines, $\Delta v_{90}$, namely, the velocity intervals encompassing 90$\%$ of the integrated optical depths. This quantity is computed as $c[\lambda(95\%) - \lambda(5\%)]/\lambda_0$, where $\lambda(5\%)$ and $\lambda(95\%)$ are the wavelengths corresponding to the 5th\ and 95th percentiles of the apparent optical depth distribution, respectively, and $\lambda_0$ is the first moment of this distribution, as defined by \cite{Ledoux2006}. 

One of the advantages of the DLA absorption spectrum is that it is produced by low-ionisation atoms dominated by galactic-scale motions governed by gravity (unlike highly ionised atoms which can make up the hot ejected gas). Therefore, these spectra are well suited for determining the dynamical velocity, which can be a proxy for their mass \citep[][]{Ledoux2006, Prochaska2008, Christensen2014, Arabsalmani2018}. Thus, by measuring $\Delta v_{90}$ in DLAs,  we can get an insight in a statistical sense on the mass of the galaxies responsible for the absorption lines. In this work, we adopted the values of $\Delta v_{90}$ from \cite{Ledoux2006, Herbert-Fort2006, Noterdaeme2008, Jorgenson2010, Rafelski2012, Neeleman2013, Berg2015} (see Table. \ref{tab:met_golden}).

\section{Method}
\label{sec:method}

The major challenge in properly determining the total (gas + dust) element abundances in the ISM is to correct account for the depletion by dust. To measure the total ISM metallicity, it is highly important to estimate the fraction of atoms of different elements withdrawn from the gas phase and incorporated into dust grains. 

\citet{DeCia2016} have developed an effective method to characterise the dust depletion in the ISM of the MW and QSO-DLAs based on the analysis of the relative abundances, that allows to derive the total (gas + dust) metallicity.
Based on this, \citet{DeCia2021} have built the so-called "relative" method to derive the total metallicty in the neutral ISM in the MW. The total abundance [X/H]$_{\rm tot}$ can be derived as follows:
\begin{equation}
\label{eq:Mtot}
   {\rm [M/H]_{tot} = [X/H] - \delta_X}
,\end{equation}
where [X/H] is the observed abundance of element X, and ${\rm \delta_X}$ is its dust depletion. The 'minus' sign in the formula \ref{eq:Mtot} is due to the fact that the value of ${\rm \delta_X}$ is negative in its classical notation \citep[e.g.][]{DeCia2016}. 

The method is aimed at deriving the total metallicity [M/H]$_{\rm tot}$ and overall strength of depletion [Zn/Fe]$_{\rm fit}$ from the distribution of element abundances [X/H] depending on their refractory index, $B2_X$.  The latter represents how strongly an element is incorporated into the dust. This linear dependence is defined as follows:
\begin{equation}
\label{eq:y}
y = a + bx
,\end{equation}
where
\begin{flalign}
x & = B2_X ,\\
y &= {\rm log}\,N(X) - {\rm log}\,N(H) - X_\odot + 12 - A2_X \sim {\rm [X/H]} ,\\
a &= {\rm [M/H]_{tot}} ,\\
b &= {\rm [Zn/Fe]_{fit}}.
\label{eq:linear_fit_params}
\end{flalign}

The coefficient $A2_X$ is the normalisation of the depletion sequences of each element after subtracting the correction for nucleosynthesis effects such as $\alpha$-element enhancement or Mn underabundance. In general it could be assumed to be zero, implying no dust depletion at [Zn/Fe]=0 \citep[][]{DeCia2016}.
The depletion factor [Zn/Fe]$_{\rm fit}$ represents the overall strength of depletion and it is derived from several different metals. \footnote{Despite using all available metals, [Zn/Fe] still has a somewhat privileged role in the analysis, i.e. the relations of ${\rm \delta_X}$ with [Zn/Fe] \citep[][]{Konstantopoulou2022}.}

The $A2_X$ and $B2_X$ coefficient have been adopted from \citet{Konstantopoulou2022}. To derive these coefficients, these authors used a large collection of column density measurements of 18 metals with different refractory properties located in various environments such as MW, the Magellanic Clouds, and DLAs.

A few remarks should be made. First, due to their nucleosynthesis peculiarities, some elements may have systematic deviations from abundance patterns that we expect from depletion effects. The best-known and most prominent examples are $\alpha$-element enhancement and Mn underabundance \citep[e.g. ][]{McWilliam1997, Mishenina2015}. 

Second, $\delta_X$ is a linear function of the overall strength of depletion characterised by the parameter [Zn/Fe]$_{\rm fit}$ in the relative method. The difference from the observed parameter [Zn/Fe] is that [Zn/Fe]$_{\rm fit}$ is derived from the abundances of all available metals or their subset.

\begin{figure*}[h!]
 \includegraphics[width=0.49\textwidth]{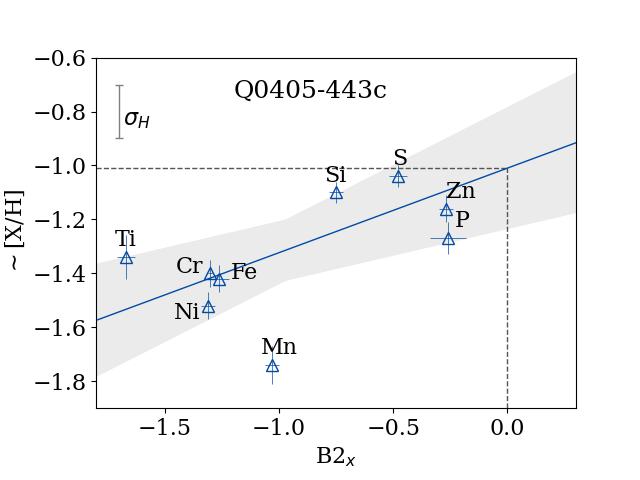} 
 \includegraphics[width=0.49\textwidth]{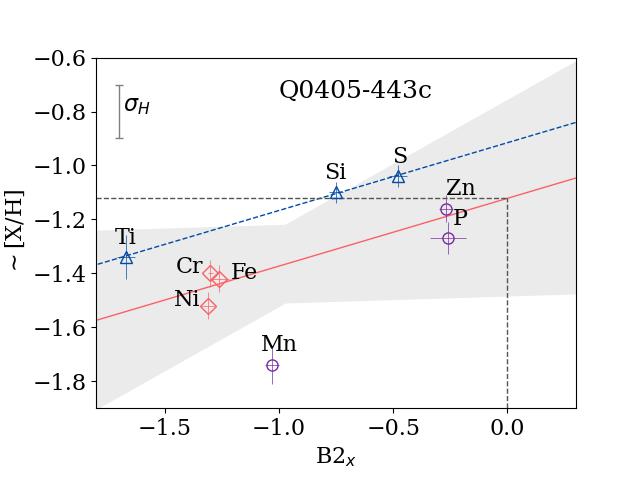} 
\caption{ Linear fitting over the entire data set of [X/H] (left panel). The elements (right panel) are divided into three groups: $\alpha$-elements (blue triangles), Fe-group elements (red diamonds), and other metals (purple circles). The blue dashed line shows linear fitting over solely the $\alpha$-element abundances, the red solid line is the blue dashed line shifted to the value averaged over Fe-group elements. The gray filled areas show 3$\sigma$ confidence intervals. The total metallicity [M/H]$_{\rm tot}$ is shown by intersection of gray dashed lines.}
\label{fig:Q0405}
\end{figure*}

Here, we consider the application of the method on the example of the QSO-DLA source Q0405--443c (see Fig. \ref{fig:Q0405}). The left panel shows the linear fit to all the element abundances. For the linear fit, we took into account error bars on both axes and did not include elements for which only the limits have been measured.  The values of [X/H] have a fairly large scatter relative to the best fit curve, which could not be explained by the 3$\sigma$ confidence interval (shown as the gray area in Fig. 2). This scatter is likely caused by the nucleosynthesis peculiarities of $\alpha$-elements and Mn.
This becomes evident from the right panel of Fig.~\ref{fig:Q0405}, where the elements have been divided into three subsets:  $\alpha$-elements (Ti, Si, S), elements of the iron group (Fe, Ni, Cr) and others. We do not include Zn to the iron group since its abundances does not always follow those of Fe, Ni and Cr, especially in dwarf galaxies \citep[see ][]{Hirai2018}. 

Despite small differences in the nucleosynthetic origin of various $\alpha$-elements \citep[see Sect. \ref{sec:intro}, ][]{Tolstoy2009}, their abundance pattern is fairly uniform, unlike other metals \citep[][see also the discussion in Sect. \ref{sec:alpha-elements}]{Prantzos2005}. It is important to separate $\alpha$-elements from other metals for the further analysis. With this assumption, the best way to derive the slope [Zn/Fe]${\rm _{fit}}$ of the linear fit to the abundance pattern is from the total amount of $\alpha$-element measurements available. 

For the correct application of this method, it is important to have an abundance measurement of titanium and at least one other $\alpha$-element. This provides sufficient dynamical range in the $x$-axis for a more confident fit because titanium is the most refractory among other $\alpha$-elements.
In six cases among the golden sample systems, the preferred slope of the linear fit to the abundance patterns would be negative. However, a negative slope is not physical because dust depletion only removes metals from the gas-phase. We limited the possible slopes to be non-negative, using the same methodology in \citet{DeCia2024}.

To obtain the total metallicity of the system, we shifted the linear fit obtained solely from the $\alpha$-elements to the weighted mean of the Fe, Cr, and Ni abundances, and derived the y-intercept. This is shown in Fig. \ref{fig:Q0405} with a red line. The value of the shifted straight line at $B2_X$=0 is the total metallicity [M/H]${\rm _{tot}}$ of the system. In this example, for Q0405-443c, we derived [M/H]${\rm _{tot}}$ = $-$1.12$\pm$0.12. 

 We derived the uncertainty on the slope [Zn/Fe]${\rm _{fit}}$ from the Python procedure we used for the linear fit to the abundance patterns. To calculate the uncertainty of the normalisation [M/H]${\rm _{tot}}$, we took into account the uncertainty, $\sigma_{\alpha_{\rm fit}}$, derived from the fitting procedure, the uncertainty of the weighted mean of the Fe-group element abundances, $\sigma_{\rm  theFe}$, and the uncertainty of hydrogen, $\sigma_{\rm H_{\rm tot}}$ as follows:
\begin{equation}
    \sigma_{\rm [M/H]_{tot}} = \sqrt{\sigma^2_{\alpha_{\rm fit}} + \sigma^2_{\rm Fe} + \sigma^2_{\rm H_{\rm tot}}}
.\end{equation}

In addition to the metallicity and amount of dust depletion, we can further study any peculiarities due to nucleosynthesis by characterising any deviations of the observed abundances from the main trend (red curve). We derived the relative abundances of different elements [X/H]${\rm _{nucl}}$ due to nucleosynthesis, after correcting for dust depletion as follows:

\begin{equation}
{\rm [X/H]_{nucl} = [X/H] - y(B2_X)}
.\end{equation}
Hence,  we obtain:
\begin{equation}
{\rm [X/Fe]_{nucl} = [X/H]_{nucl} - [Fe/H] _{nucl}}
.\end{equation}
\section{Results and discussion}
\label{sec:results}

We analysed the abundance patterns of 110 DLAs, in particular, 24 DLAs in the golden sample. In most cases, we find a satisfactory fit to the data. These are shown in Figs. \ref{fig:XH_B2X_golden1} and \ref{fig:XH_B2X_non-golden1}. Tables \ref{tab:met_golden} and \ref{tab:abund_golden} contain the resulting data for the golden sample: total (gas+dust) metallicities, [M/H]\tot, the overall strength of dust depletion, [Zn/Fe]\fit, and the relative abundances due to nucleosynthesis, [X/Fe]$_{\rm nucl}$, which are the deviations from the abundance patterns after taking dust depletion into account.

\begin{figure}
\centering
\resizebox{\hsize}{!}
   {\includegraphics{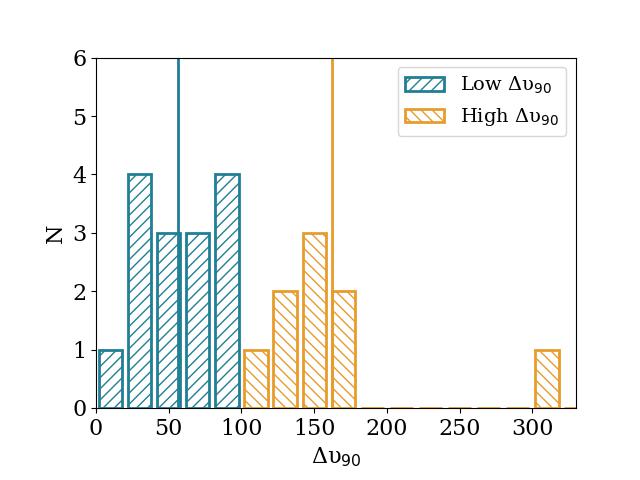}}
  \caption{Distribution of the golden sample by velocity widths.}
\label{fig:dv90-amount}
\end{figure}

We divided the golden sample into two parts by velocity widths: $\Delta v_{90} < 100$\kms\, and $\Delta v_{90} \geq 100$\kms\, which we refer to as low- and high-$\Delta v_{90}$ subsamples, respectively. Figure~\ref{fig:dv90-amount} shows the distribution of QSO-DLAs by $\Delta v_{90}$. The dividing point has been chosen in such a way that the two subsamples roughly contains a comparable number of systems: there are 15 and 9 QSO-DLAs in low- and high-$\Delta v_{90}$ subsamples, respectively. Using 70 DLA or sub-DLA systems \citet{Ledoux2006} have shown a correlation between metallicity and velocity widths which is probably the consequence of an underlying mass-metallicity relation \citep[][]{Christensen2014}. Thus, the low-\dvn\, sub-sample is expected to have overall lower galaxy masses than the high-\dvn\, sub-sample. According to \citet{Arabsalmani2018}, the threshold velocity width of 100 \kms\, corresponds to the stellar mass of the galaxy $M^*$ of about 10$^9$M$_\odot$.

\begin{figure}
\centering
\resizebox{\hsize}{!}
   {\includegraphics{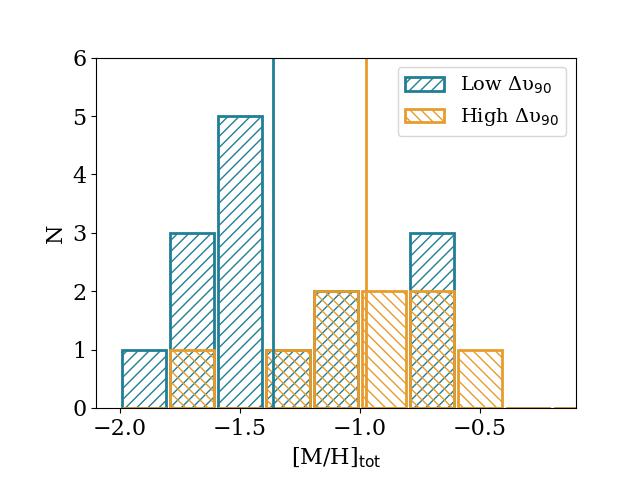}}
  \caption{Distribution of the golden sample by metallicity [M/H] for low (shown by blue) and high (orange) $\Delta v_{90}$ subsamples.}
\label{fig:met-amount}
\end{figure}

Figure~\ref{fig:met-amount} shows the distribution of the golden sample by the total metallicity. We see that [M/H]\tot\, varies from $-$2.0 to $-$0.5. The metallicity ranges for the two subsamples mostly overlap, although the distributions are different: vertical lines show the weighted mean which is $-$1.36 and $-$0.97 for the low- and high-$\Delta v_{90}$ subsamples, respectively.

\begin{figure}
\centering
\resizebox{\hsize}{!}
   {\includegraphics{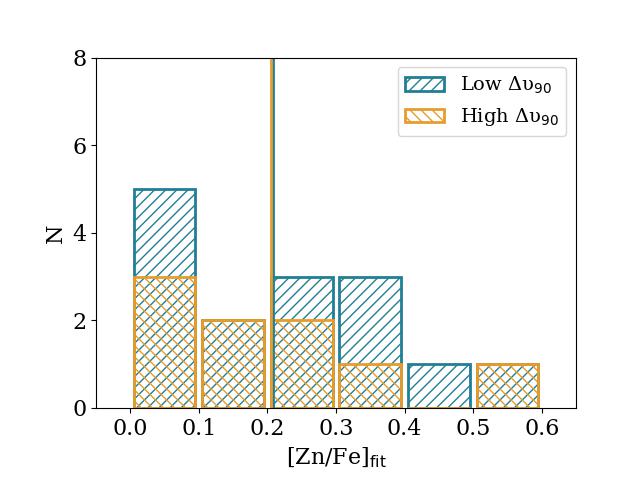}}
  \caption{Distribution of the golden sample by dust depletion [Zn/Fe]$_{\rm fit}$ for low (shown by blue) and high (orange) $\Delta v_{90}$ subsamples. }
\label{fig:slope-amount}
\end{figure}

The strength of dust depletion [Zn/Fe]\fit\, shown in Fig.~\ref{fig:slope-amount} is in the range from 0.0 to 0.55, with the two subsamples displaying only a very small systematic difference in distribution of 0.01 (see vertical lines in Fig.~\ref{fig:slope-amount} that show the weighted mean).

\begin{figure}
\centering
\resizebox{\hsize}{!}
   {\includegraphics{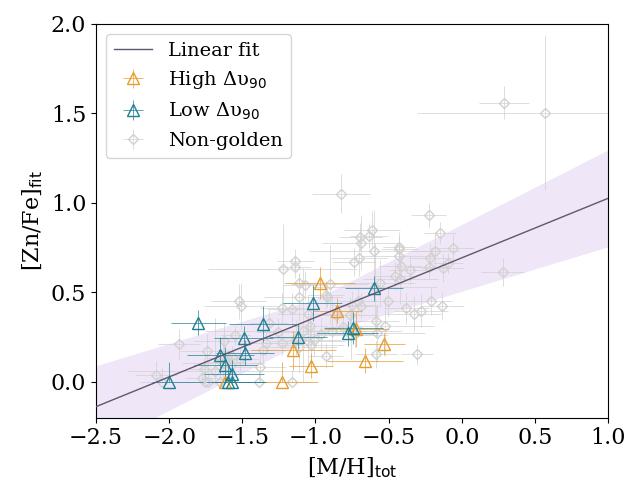}}
  \caption{Relation between the total metallicity, [M/H]\tot\, and the strength of dust depletion, [Zn/Fe]\fit, in the neutral ISM of DLAs. Large blue and orange triangles show the data for the golden sample with low- and high-\dvn, respectively. The small gray diamonds are the data for the non-golden sample. The solid line shows a linear fit to all the data including uncertainties on both axes, [Zn/Fe]\fit\,= 0.33$\pm$0.03$\times$[M/H]\tot\, + 0.69$\pm$0.03 . The shaded area represents the 3$\sigma$ confidence interval.}
\label{fig:M-H_Zn-Fe}
\end{figure}

Figure \ref{fig:M-H_Zn-Fe} displays a broad correlation between the total metallicity and the strength of dust depletion with quite a large scatter. Similar relations have been derived by other authors \citep[][]{DeCia2024, Noterdaeme2008, Ledoux2002}. Hence, the amount of dust in DLAs overall depends on the gas metallicity. This is in agreement with an assumption that significant amount of dust is built through grain-growth in the ISM depending on the gas metallicity, density, and temperature  \citep[][]{Dwek2016, Mattsson2014, DeCia2013}. From Fig. \ref{fig:M-H_Zn-Fe}, it might appear that low-$\Delta v_{90}$ systems show larger [Zn/Fe]$_{\rm fit}$ compared to high-$\Delta v_{90}$ galaxies. 
In fact, this is a shift in metallicity (see Fig. \ref{fig:met-amount}), while [Zn/Fe]$_{\rm fit}$ values are similarly distributed in low- and high-$\Delta v_{90}$ galaxies, as shown in Fig. \ref{fig:slope-amount}. The data for the MW do not strongly follow the dust-metallicity relation showing steeper slope than DLAs. The possible cause of this is that the MW disk has higher pressure and higher density of colder gas compared to more diffuse warm neutral medium in DLAs \citep[see][]{DeCia2024}. As a result, dust is expected to grow more easily in the MW.

\begin{figure}
   \includegraphics[width=1.0\linewidth]{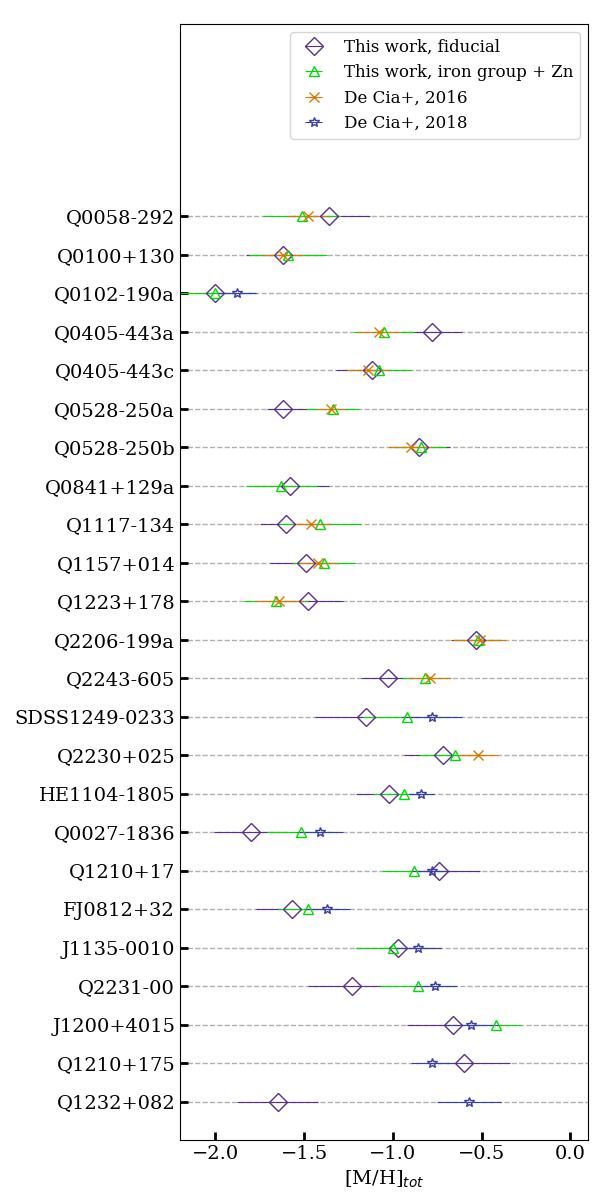}
  \caption{Total metallicity [M/H]${\rm _{tot}}$ in the ISM of QSO-DLAs from the golden sample. Large purple diamonds are the values  derived from solely $\alpha$-elements (this work, the basic result). Small green triangles show the values derived from fitting the iron-group elements + Zn. Small orange crosses and small blue stars are the determinations given by \citet{DeCia2016} and \citet{DeCia2018}, respectively.}
\label{fig:M-H_tot_comp}
\end{figure}

 We compared our most solid derivation of the total metallicity, [M/H]$_{\rm tot}$, for the golden sample with alternative ways of estimating the metallicity, to test their robustness. First, we compared our results to the total metallicity based on the fit of the abundance pattern of only the Fe-group elements and Zn (green triangles in Fig. \ref{fig:M-H_tot_comp}). This reveals that in most cases, the estimates based only on the Fe-group are slightly lower than the fiducial value, on average by 0.06 dex. The values of total metallicity determined by \citet{DeCia2016} and \citet{DeCia2018} (yellow crosses and blue stars in Fig. \ref{fig:M-H_tot_comp})  generally demonstrae a greater similarity to the metallicity we determined using only the Fe-group elements and Zn. This reflects the fact the methodologies of \citet{DeCia2016} and \citet{DeCia2018} based their results more heavily on Fe-group elements and Zn. In this work, however, the methodology is refined to separate the effects of metallicity, dust depletion, and nucleosynthesis, and thus more robust. One outstanding example is the case of Q1232+082, for which we measure a total metallicity, which is ~1 dex lower than the estimate of \citet{DeCia2016}. This is due to high $\alpha$-element enhancement (0.5 dex, see Fig. \ref{fig:XH_B2X_golden1}) in this case, combined with the absence of measurements of Ti in the work of \citet{DeCia2016}. As a result, the slope of the linear fitting is steeper which leads to the zero intercept at a higher value of [M/H]$_{\rm tot}$. 

\renewcommand{\arraystretch}{1.2}%
\begin{table*}[]
    \centering
    \caption{Collection of data for QSO-DLAs in the golden sample. The values of logN(HI) were taken from \citet{Konstantopoulou2022} or \citet{DeCia2016}.}
    \begin{tabular}{lcccccc}
\hline
QSO & $z_{abs}$ & logN(HI) & [M/H]$_{\rm tot}$ & [Zn/Fe]$_{\rm fit}$ & $\Delta v_{90}$ & Ref. \\
\hline
Q0058$-$292     & 2.671 & 21.10 $\pm$ 0.10 & $-$1.36 $\pm$ 0.23 & 0.32 $\pm$ 0.10          & 34$^a$  & 1,2\\ 
Q0100+130     & 2.309 & 21.35 $\pm$ 0.08 & $-$1.62 $^{+0.22}_{-0.21}$ & 0.09$^{+0.10}_{-0.09}$   & 37$^a$ & 1,2 \\ 
Q0102$-$190a    & 2.370 & 21.00 $\pm$ 0.08 & $-$2.00 $^{+0.23}_{-0.12}$ & 0.00$^{+0.11}_{-0.00}$   & 17$^a$  & 1,2\\ 
Q0405$-$443a    & 1.913 & 20.80 $\pm$ 0.10 & $-$0.78 $\pm$ 0.21 & 0.27 $\pm$ 0.07          & 98$^a$  & 1,2\\ 
Q0405$-$443c    & 2.595 & 21.05 $\pm$ 0.10 & $-$1.12 $\pm$ 0.20 & 0.25 $\pm$ 0.07          & 79$^a$  & 1,2\\ 
Q0528$-$250a    & 2.141 & 20.98 $\pm$ 0.05 & $-$1.62 $^{+0.17}_{-0.09}$ & 0.00$^{+0.08}_{-0.00}$   & 105$^a$  & 1,2\\ 
Q0528$-$250b    & 2.811 & 21.35 $\pm$ 0.07 & $-$0.85 $\pm$ 0.17 & 0.39 $\pm$ 0.07          & 304$^a$  & 1,2\\ 
Q0841+129a    & 1.864 & 21.00 $\pm$ 0.10 & $-$1.58 $^{+0.22}_{-0.19}$ & 0.04$^{+0.08}_{-0.04}$   & 32$^a$  & 1,2\\ 
Q1117$-$134     & 3.350 & 20.95 $\pm$ 0.10 & $-$1.60 $^{+0.25}_{-0.15}$& 0.00$^{+0.10}_{-0.00}$    & 44$^a$  & 1,2\\ 
Q1157+014     & 1.944 & 21.80 $\pm$ 0.10 & $-$1.49 $\pm$ 0.20 & 0.24 $\pm$ 0.07          & 89$^a$  & 1,2\\ 
Q1223+178     & 2.466 & 21.40 $\pm$ 0.10 & $-$1.48 $\pm$ 0.20 & 0.16 $\pm$ 0.07          & 91$^a$  & 1,2\\ 
Q2206$-$199a    & 1.921 & 20.67 $\pm$ 0.05 & $-$0.53 $\pm$ 0.14 & 0.21 $\pm$ 0.06          & 136$^a$  & 1,2\\ 
Q2243$-$605     & 2.331 & 20.65 $\pm$ 0.05 & $-$1.03 $\pm$ 0.15 & 0.09 $\pm$ 0.07          & 173$^a$  & 1,2\\ 
SDSS1249$-$0233 & 1.781 & 21.45 $\pm$ 0.15 & $-$1.15 $\pm$ 0.29 & 0.18 $\pm$ 0.11          & 152$^b$  & 3\\ 
Q2230+025     & 1.864 & 20.83 $\pm$ 0.05 & $-$0.72 $\pm$ 0.22 & 0.29 $\pm$ 0.10          & 148$^a$ & 3\\ 
HE1104$-$1805   & 1.662 & 20.85 $\pm$ 0.01 & $-$1.02 $\pm$ 0.20 & 0.44 $\pm$ 0.10          & 50$^c$ & 3\\ 
Q0027$-$1836    & 2.402 & 21.75 $\pm$ 0.10 & $-$1.80 $\pm$ 0.19 & 0.33 $\pm$ 0.07          & 44$^d$ & 3\\ 
Q1210+17      & 1.892 & 20.63 $\pm$ 0.08 & $-$0.74 $\pm$ 0.20 & 0.30 $\pm$ 0.09          & 62$^a$ & 3\\ 
FJ0812+32     & 2.067 & 21.00 $\pm$ 0.10 & $-$1.57 $^{+0.23}_{-0.14}$ & 0.00$^{+0.09}_{-0.00}$   & 26$^{e,*}$ & 3\\ 
J1135$-$0010    & 2.207 & 22.05 $\pm$ 0.10 & $-$0.97 $\pm$ 0.24 & 0.55 $\pm$ 0.09          & 168$^f$ & 3\\ 
Q2231$-$00      & 2.066 & 20.53 $\pm$ 0.08 & $-$1.23 $^{+0.25}_{-0.14}$ & 0.00$^{+0.11}_{-0.00}$   & 145$^a$ & 3\\ 
J1200+4015    & 3.220 & 20.65 $\pm$ 0.15 & $-$0.66 $\pm$ 0.26 & 0.12 $\pm$ 0.07          & 127$^g$ & 3\\ 
Q1210+175     & 1.892 & 20.70 $\pm$ 0.08 & $-$0.60 $\pm$ 0.20 & 0.52 $\pm$ 0.07          & 62$^a$ & $\star$,1\\ 
Q1232+082     & 2.338 & 20.90 $\pm$ 0.08 & $-$1.65 $\pm$ 0.23 & 0.15 $\pm$ 0.10          & 85$^a$ & $\star$,1\\ 
\hline 
\end{tabular}
\begin{tablenotes}
\item[]{
\footnotesize
{\bf Notes:} The values of $\Delta v_{90}$ have been taken from 
             $^a$ -- \cite{Ledoux2006}; 
             $^b$ -- \cite{Herbert-Fort2006};
             $^c$ -- \cite{Neeleman2013};
             $^d$ -- \cite{Noterdaeme2008};
             $^e$ -- \cite{Jorgenson2010};
             $^f$ -- \cite{Christensen2019};
             $^g$ -- \cite{Berg2015}.
 Symbol $^*$ designates a very poorly studied system, for which it is not clear what the extent of the metal-line profile is. Only lower limit has been determined.  Column densities are taken from (1) \protect\cite{DeCia2016}, (2) \protect\cite{Konstantopoulou2022}, (3) \protect\cite{Berg2015} \protect\citep[corrected by][to the newest oscillator strengths, see Sec. \ref{sec:column_densities}]{Konstantopoulou2022}, and ($\star$) this work.}
 \end{tablenotes}
 \label{tab:met_golden}
\end{table*}

\subsection{$\alpha$-elements}
\label{sec:alpha-elements}

The total $\alpha$-element abundances [$\alpha$/Fe]\nucl\, for each object has been obtained by calculating an average weighted by the uncertainties over all available individual $\alpha$-element (Ti, Si, S, Mg, and O) values [X/Fe]\nucl.
Looking at the $\alpha$-element enhancement [$\alpha$/Fe], as calculated from the fitting described in Sect. \ref{sec:method}, for our entire sample does not yield any obvious relation as seen for individual galaxies in Fig. \ref{fig:alpha-met-gal}. This is expected as the DLA population probes a large range of different galaxies \citep[e.g.][]{Fynbo2008, Krogager2017} that will have heterogeneous enrichment histories \citep[see also][]{Dvorkin2015}. However, when splitting our sample into bins of \dvn, a correlation is apparent. 

\begin{figure*}
   \includegraphics[width=0.49\linewidth]{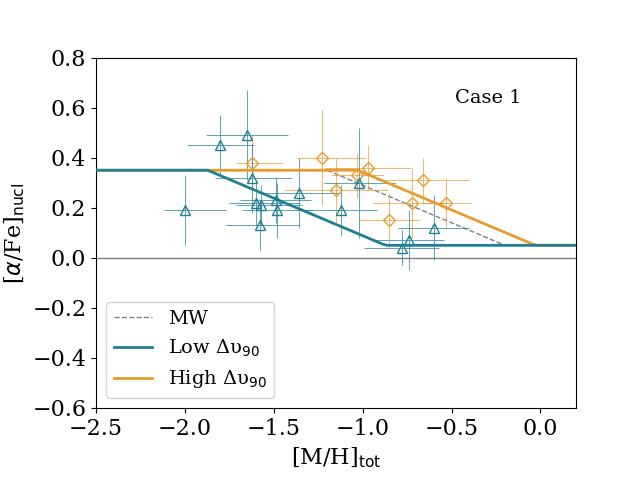}
   \includegraphics[width=0.49\linewidth]{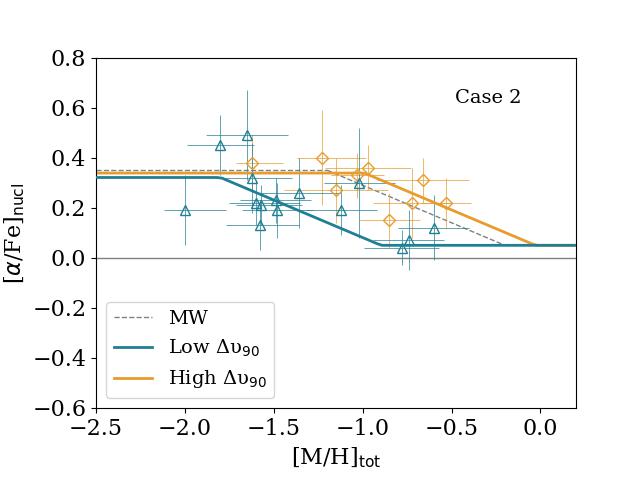}
   \includegraphics[width=0.49\linewidth]{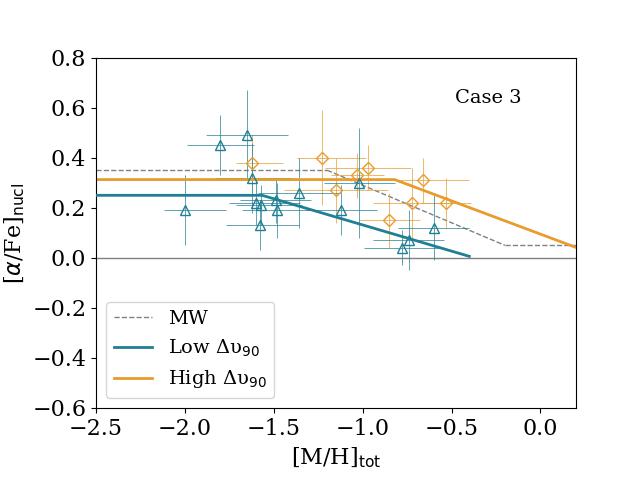}
   \includegraphics[width=0.49\linewidth]{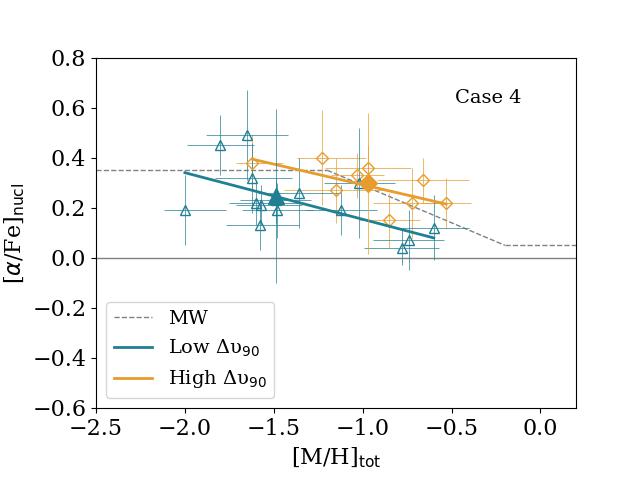}   
  \caption{$\alpha$-elements abundances for low-$\Delta v_{90}$ (blue open triangles) and high $\Delta v_{90}$ (orange open diamonds) QSO-DLA systems. Case 1:  Data have been approximated by three-piecewise function: high-$\alpha$ plateau + decline + low-$\alpha$ plateau. The five parameters of the function are: levels of high- and low-$\alpha$ plateaus, and the slope and positions of the high- and low-$\alpha$ knees. During the fitting, only the position of the high-$\alpha$ knee is considered to be free, while the other parameters are fixed and set to values given for the MW. The gray dashed curve shows behaviour for the MW as determined by \protect\cite{McWilliam1997}. Case 2: The level of the high-$\alpha$ plateau and position of the high-$\alpha$ knee are allowed to vary.  Case 3: Fitting the data with a two-piecewise function. Case 4: Fitting the data with a linear function. Two filled points at the lower right panel indicate median values in [M/H]\tot\, and average values in [$\alpha$/Fe]\nucl\, show the systematic difference between the two subsamples. All the parameters and their uncertainties are shown in Table \ref{tab:alpha-params}.} 
\label{fig:alpha-met-DLAs}
\end{figure*}

In Fig. \ref{fig:alpha-met-DLAs}, we show [$\alpha$/Fe] for our golden sample as a function of total metallicity in the low- and high-\dvn\, subsamples (as blue and orange points, respectively). Since $\Delta v_{90}$ is a proxy of galactic mass, it is reasonable to expect some difference between the high-$\alpha$ knee positions for these two groups of systems (see discussion in Section \ref{sec:alpha-elements}). It is evident that the orange points tend to cluster at larger metallicities than the blue points, as expected from the correlation between \dvn\, and metallicity \citep[][]{Ledoux2006}. Yet, the two subsamples also seem to follow a similar anti-correlation between [$\alpha$/Fe] versus [M/H]\tot. This is illustrated in the simplest case by a linear fit to the two subsamples (Case 4). For both subsamples, we find consistent slopes of $-$0.19$\pm$0.15 and $-$0.16$\pm$0.22 for the low- and high-\dvn\, subsamples, respectively, with an offset in metallicity of $\sim$0.7 dex. 

Both of the inferred slopes are slightly lower than the slope inferred for the MW \citep[$-$0.3, see][]{McWilliam1997}, yet consistent within the considerable uncertainties that are dominated by scatter in the data. Upon further inspection, there tends to be a slight flattening of the two relations: the relation for the low-\dvn\, subsample appears reach a low-alpha plateau at higher metallicities, $\gtrsim-1$, whereas the high-\dvn\, subsample tends towards a high-$\alpha$ plateau at low metallicity, $\lesssim-1.2$. This behaviour follows the expectation from the MW and local dwarf galaxies \citep[see Fig. \ref{fig:alpha-met-gal} and ][]{McWilliam1997}. The presence of a plateau in the data would indeed bias the slopes towards lower values. We consider four possibilities for fitting the data, which are shown in Fig.~\ref{fig:alpha-met-DLAs} and reported in Table \ref{tab:alpha-params}:
\begin{itemize}
  \item Case 1: Motivated by the behaviour of [$\alpha$/Fe] in the MW and local dwarf galaxies, we fit our data with a three-piecewise function (high-$\alpha$ plateau + decline + low-$\alpha$ plateau) with five parameters. Due to the strong spread of values and a small number of points, an attempt to set all parameters as free leads to a large uncertainty in fitting.  As a first guess,  we left the position of the $\alpha$-element knee free to vary and we fixed the other parameters according to the values defined for the MW by \cite{McWilliam1997}: the level of high-$\alpha$ plateau [$\alpha$/Fe]$_{\rm nucl}$ is 0.35, the slope is $-$0.3, and the level of low-$\alpha$ plateau is 0.05 (see Table \ref{tab:alpha-params}).
  \item Case 2: We fixed the slope and the low-$\alpha$ plateau at the values of \citet{McWilliam1997} and fit the data with the three-piecewise functions to find the high-$\alpha$ knee positions and the high-$\alpha$ plateau levels. 
  \item Case 3: Since the paucity of observational data of [$\alpha$/Fe]$_{\rm nucl}$ hampers our capability to constrain all the five parameters of a three-piecewise function, we tested the results by applying a two-piecewise function (high-$\alpha$ plateau + decline) with all three parameters remaining free. 
  \item Case 4: Fitting the data with a linear function of the form: [$\alpha$/Fe] $= \alpha_0$ $+$ slope $\times$ [M/H]\tot.
\end{itemize}

Because of the large scatter of points, the values of all the parameters vary from case to case, but they are consistent within the error bars (see Table \ref{tab:alpha-params}). For the high-\dvn\, subsample, we can clearly determine the high-$\alpha$ plateau, high-$\alpha$ knee, and slope. However, the low-alpha knee and low-alpha plateau cannot be constrained due to the lack of points at metallicities lower than $-$0.5 dex. For the low-\dvn\, subsample, the plateau seems to be less clearly constrained, unlike the high-\dvn\, one (see also Fig. \ref{fig:S-Si-Ti-met-DLAs-add} in the appendix to assess the variations between different $\alpha$-elements). Overall, all the models show systematic differences between the low- and high-\dvn\, subsamples in the [$\alpha$/Fe]\nucl\, -- [M/H]\tot\, plane. These are probably due to different chemical evolution phases and properties (e.g. SFR and SFH) of galaxies with different stellar masses.

To compare these four cases, we computed the reduced $\chi^2$ ($\chi^2_\nu$) values and report them in Table~\ref{tab:alpha-params}. From the values, it was impossible to choose a preferred model because the difference in $\chi^2_\nu$ should be larger than $\sim$1 (or $\sim$2$\sigma$) to state that a given model is better over another.

\begin{table}[]
    \centering
    \caption{Parameters of the piecewise function fitting $\alpha$-element enhancement in the low- and high-$\Delta v_{90}$ subsamples. Case 1 is shown in Fig. \ref{fig:alpha-met-DLAs}.}
    \begin{tabular}{lcccc}
\hline
Case & $\alpha$ knee          & Plateau level &  Slope  & $\chi^2_\nu$      \\
     & [M/H]$_{\rm tot}$, dex &  [$\alpha$/Fe]$_{\rm nucl}$, dex & &\\
\hline
\multicolumn{5}{c}{High $\Delta v_{90}$} \\
\hline
1 & $-$1.03$\pm$0.15   & 0.35$^*$       & $-$0.30$^*$    &   0.39 \\
2 & $-$1.01$\pm$0.28   & 0.34$\pm0.05$  & $-$0.30$^*$   &    0.45  \\
 3 & $-$0.82$\pm$0.66   &  0.31$\pm$0.04  &  $-$0.26$\pm$0.81  &  0.60\\
4 &   ...           &  ...            &  $-$0.16$\pm$0.22  & 0.37 \\
\hline
\hline
\multicolumn{5}{c}{Low $\Delta v_{90}$} \\
\hline
1 & $-$1.87$\pm0.11$   & 0.35$^*$       & $-$0.30$^*$    &    0.58  \\
2 & $-$1.80$\pm0.32$   & 0.32$\pm0.09$  & $-$0.30$^*$    &    0.60  \\
3 & $-$1.56$\pm$0.38   &  0.25$\pm$0.04  &  $-$0.21$\pm$0.11  & 0.71\\
4 &    ...            &  ...            &  $-$0.19$\pm$0.15  & 0.62 \\
\hline
\multicolumn{4}{l}{$^*$ -- fixed parameters} \\
\end{tabular}
    \label{tab:alpha-params}
\end{table}

\begin{figure}
  \includegraphics[width=1.0\linewidth]{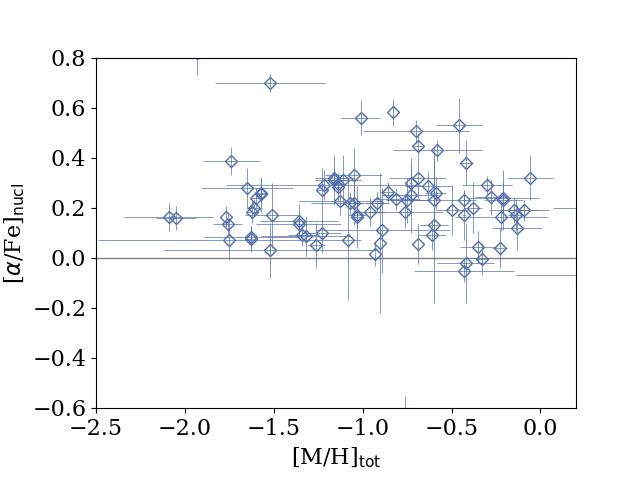}
  \caption{$\alpha$-element overabundance with respect to Fe as a function of the total metallicity for the non-golden sample. Two points are out of the range of the figure at ([M/H]\tot, [$\alpha$/Fe]) = ($-$1.93$\pm$0.15, 0.82$\pm$0.09) and ($-$0.76$\pm$0.13, $-$0.63$\pm$0.08).}
\label{fig:alpha-all-DLAs}
\end{figure}

For a comparison with the results shown above, we applied the same fitting technique to the APOGEE stellar data in nearby dwarf galaxies. This is shown in Figs \ref{fig:alpha-met-gal} and \ref{fig:alpha-galaxies}. It is only for Sgr that the data allow for  full three-piecewise function to be  traced and the high-$\alpha$ knee is at [M/H]=$-$1.92$\pm$0.05. From the data, it is impossible to determine the position of the high-$\alpha$ knee for the SMC, LMC, and Fnx galaxies. \citet{Nidever2020} constrained it to be at [Fe/H]$\lesssim-$2.2 for both LMC and SMC, despite the fact that LMS is 10 times more massive than SMC. The abundances derived by \citet{Van-der-Swaelmen2013} from high-resolution spectra at VLT with the FLAMES/GIRAFFE multifibre spectrograph show higher $\alpha$-element enhancement for the LMC. Also, there is some hint that the knee is at higher metallicity of $\approx-$1.5.

\citet{Hendricks2014} reported that the high-$\alpha$ knee for Fnx dwarf galaxy is at [Fe/H]$\approx-$1.9, which was determined from abundance measurements of Mg using VLT/GIRAFFE spectra.
The position of the knee is less clearly defined for Si and Ti, but apparently it is more clearly seen at metallicities below $-$1.8. From the APOGEE data, there are no signs of the presence of high-$\alpha$ plateau at metallicities above $-$2.0 (see Figs. \ref{fig:alpha-met-gal} and \ref{fig:alpha-galaxies}). We derived the position of the high-$\alpha$ knee for GSE to be at [M/H] of $-$1.21$\pm$0.01.

The high-$\alpha$ knee is challenging to determine even in local galaxies because it requires measuring reliable $\alpha$-element abundances for a statistically significant number of faint metal-poor stars. Furthermore, there are some discrepancies among different observations \citep[][]{Nidever2020, Hendricks2014, Van-der-Swaelmen2013} that causes additional confusion in trying to interpret the results.
Compared to GSE and Sgr, the high-$\alpha$ knee in LMC is more metal-poor \citep[][]{Nidever2020} even though LMC is more massive than the progenitors of GSE and Sgr (see Table \ref{tab:galaxies-params}). This can be explained by the differences in the early evolution of the Magellanic Clouds from that of many other local galaxies \citep[][]{Nidever2020}.
Obviously, Local Group galaxies do not fully fit the simplified picture where faint, low-mass galaxies show lower enrichment efficiencies. Their evolutionary histories depend on several factors such as the total stellar mass, amount of gas, interaction with the MW, efficiency in producing metals, and so on. However, it is not the goal of this paper to reconcile the stellar observations in nearby galaxies.

\begin{table}[]
    \centering
    \caption{ Stellar M$_*$ and dynamical M$_d$ masses of nearby galaxies or their progenitors in cases of GSE and Sgr. The data are from $^{\rm O}$ -- \protect\citet{DOnghia2016}, $^{\rm L}$ -- \protect\citet{Limberg2022}, $^{\rm W}$ -- \protect\citet{Walker2006}, $^{\rm B-C} $ -- \protect\citet{Bermejo-Climent2018}.}
    \begin{tabular}{lcc}
\hline
galaxy & M$_*$, (M$_\odot$)  & M$_d$, (M$_\odot$)  \\
\hline
SMC    & 3$\times$10$^8$, $^{\rm O}$    & 2.4$\times$10$^9$, $^{\rm O}$      \\
LMC    & 3$\times$10$^9$, $^{\rm O}$    & 1.7$\times$10$^10$, $^{\rm O}$     \\
GSE    & 1.3$\times$10$^9$, $^{\rm L} $ & ...                                \\
Sgr    & $\sim$10$^9$, $^{\rm B-C} $      & ...                                \\
Fnx    & $\sim$10$^{7}$, $^{\rm B-C} $  & 10$^8$ -- 10$^9$, $^{\rm W} $      \\
\hline
\end{tabular}
    \label{tab:galaxies-params}
\end{table}

\begin{figure}
   \includegraphics[width=1.0\linewidth]{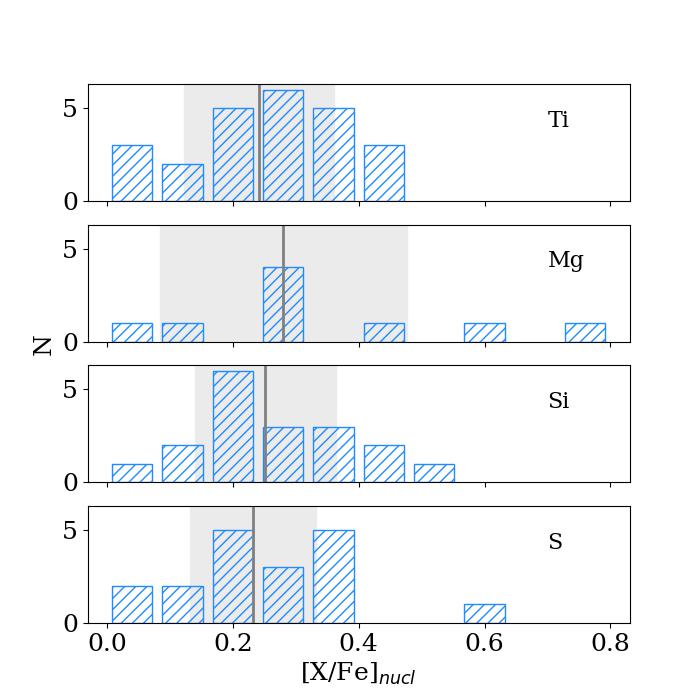}
  \caption{ Distribution of [X/Fe]$_{\rm nucl}$ for Ti, Si, S, and Mg for the golden sample. Gray vertical lines show the weighted average, while the filled areas correspond to the standard deviation.}
\label{fig:alpha_distr}
\end{figure}

Figure \ref{fig:alpha_distr} shows the distribution of [X/Fe]\nucl\, for different $\alpha$-elements in QSO-DLAs from the golden sample -- except for oxygen, for which only one measurement is available. The weighted averages of [X/Fe]$_{\rm nucl}$ (gray vertical lines in Fig. \ref{fig:alpha_distr}) vary between the elements within 0.05 dex with standard deviation (filled areas in Fig. \ref{fig:alpha_distr}) between 0.1 and 0.2. Thus, the abundance patterns of different $\alpha$-elements are fairly uniform and a splitting sample based on available elements is not expected to produce sufficiently different [$\alpha$/Fe]\nucl\, versus [M/H]\tot\, behaviour. This also can be seen from Fig. \ref{fig:S-Si-Ti-met-DLAs-add}, where we show the fitting of [X/Fe]--[M/H]\tot\, for Ti, S, and Si (if available), with three-piecewise functions. For the high-\dvn\, subsample, the high-$\alpha$ plateau  varies from [$\alpha$/Fe]\tot\,= 0.31$\pm$0.03 to 0.36$\pm$0.02, while the position of the knee is within [M/H]\tot\,= $-1.09\pm0.13$ -- $-0.80\pm0.18$. For the low-\dvn\, subsample, the parameter determination is less stable due to lack of data, especially at low metallicities. From [Si/Fe]\nucl\,, the parameters are not constrained, while from Ti and S the plateau is 0.39$\pm$0.02 and 0.25$\pm$0.02 and the knee is $-1.84\pm0.44$ and $-1.34\pm0.11$.

\subsection{Manganese}

\begin{figure}
   \includegraphics[width=1.0\linewidth]{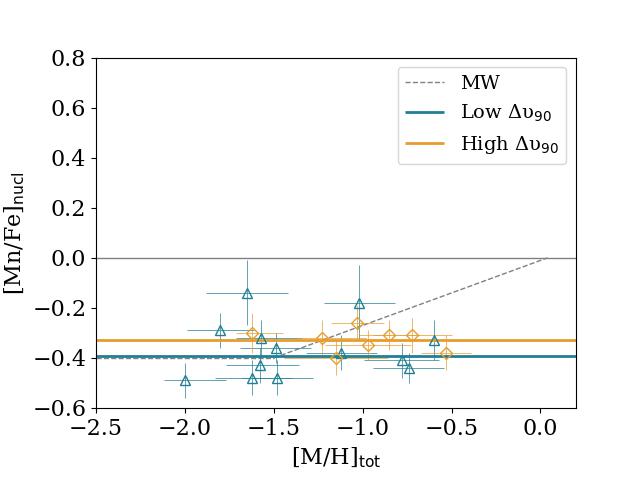}
    \includegraphics[width=1.0\linewidth]{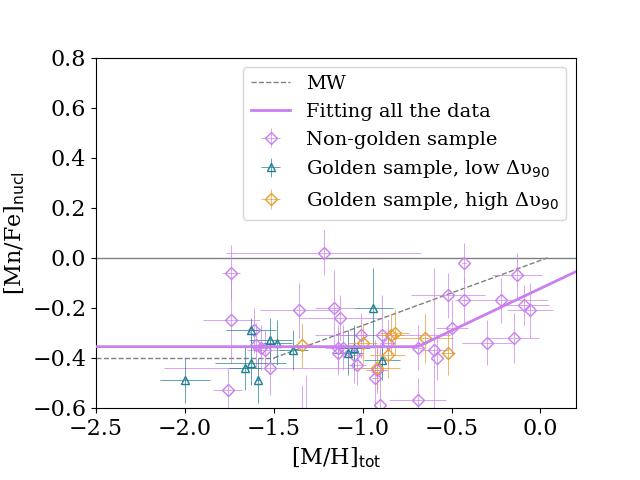}
  \caption{Manganese abundances in the ISM of DLAs. Upper panel: Mn abundances for the golden sample for low-$\Delta v_{90}$ (blue open triangles) and high-$\Delta v_{90}$ (orange open diamonds). Horizontal color lines show the weighted average values of $-$0.39$\pm$0.10 and $-$0.33$\pm$0.04 for the low- and high- $\Delta v_{90}$ subsamples, respectively. 
  Lower panel: Mn abundances for the non-golden sample (magenta open diamonds) as well as for the golden sample (symbols as on the upper panel) with [Mn/Fe]$_{\rm nucl}$ and [M/H]${\rm _{tot}}$ being derived according to the methodology used for the non-golden sample. Two points are out of the range of the figure at ([M/H]\tot, [Mn/Fe]) = ($-$1.32$\pm$0.22, $-$0.63$\pm$0.16), and ($-$1.34$\pm$0.08, $-$0.66$\pm$0.15).
  In both panels, the gray dashed curve shows typical behaviour for the MW taken from \protect\cite{Mishenina2015}.}
\label{fig:MN-DLAs}
\end{figure}

The distribution of [Mn/Fe]$_{\rm nucl}$ with metallicity is shown in Fig.~\ref{fig:MN-DLAs}. The top panels shows the results for the golden sample. There is a clear Mn underabundance for all DLAs with weighted mean values of $-$0.33$\pm$0.07 and $-$0.39$\pm$0.07 for the high- and low-$\Delta v_{90}$ subsamples, respectively. We do not see a hint of an increase of [Mn/Fe]$_{\rm nucl}$ with metallicity up to [Fe/H]$_{\rm tot}$=$-$0.53 dex. This is contrary to our expectations because it is clear from the behaviour of $\alpha$-elements in Fig. \ref{fig:alpha-met-gal} that in some DLAs, Type Ia SNa should contribute to the nucleosynthesis, which, in turn, should lead to an increase in the abundance of manganese relative to iron. In addition, observations of stellar abundances of Mn both in the MW and in nearby galaxies show increasing [Mn/Fe]$_{\rm nucl}$ toward higher metallicity starting from [Fe/H]$\sim-$1.5 dex (see gray dashed curve in Fig. \ref{fig:MN-DLAs} for the averaged behaviour in the MW).  From the analysis of elemental abundances in the neutral ISM of the SMC, \citet{DeCia2024} observed consistent [$\alpha$/Fe]$_{\rm nucl}$ enhancement and [Mn/Fe]$_{\rm nucl}$ under-abundance at an approximately constant level throughout the whole range of metallicity, which can be explained by contribution from recent core-collapse SNe.

For the non-golden sample, a hint of increase of [Mn/Fe]$_{\rm nucl}$ is seen starting from [M/H]$_{\rm tot} \approx-$0.7 dex (lower panel of Fig. \ref{fig:MN-DLAs}).  For the consistency of this comparison, the figure has been supplemented with the data for the golden sample, but in this case the [Mn/Fe]$_{\rm nucl}$ and [M/H]${\rm _{tot}}$ values has been determined according to the methodology used for the non-golden sample. We approximated the entire data set with a two-piecewise function consisting of a plateau and an increase. The best-fitting parameters are the following: the plateau is at the level of [Mn/Fe]$_{\rm nucl}=-0.35\pm$0.02, the knee position is at [M/H]${\rm _{tot}}=-0.69\pm$0.13 and the slope is 0.34$\pm$0.10.

\subsection{Phosphorus}

\begin{figure}
   \includegraphics[width=1.0\linewidth]{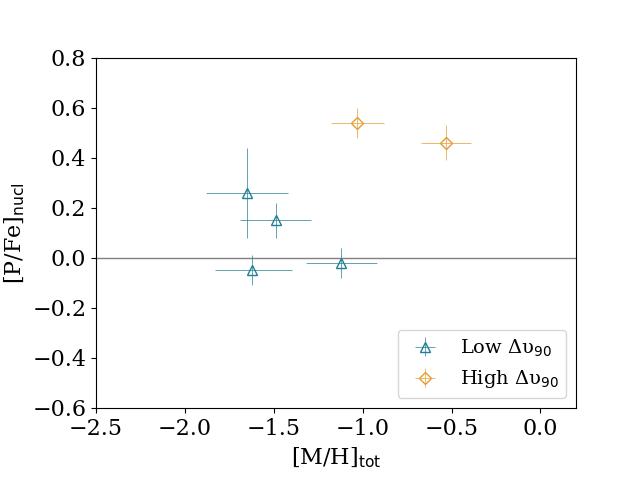}
  \caption{Phosphorus abundances in the ISM of DLAs for low-$\Delta v_{90}$ (blue open triangles) and high $\Delta v_{90}$ (orange open diamonds).}
\label{fig:P-DLAs}
\end{figure}

\citet{Konstantopoulou2023} found a most reliable value of refractory index $B2_{\rm P}$ for DLAs equal to $-$0.26$\pm$0.08, which we used in this study. For Q1232+082, we find that the column density measurement for the bluest spectral component of the PII $\lambda$1152 line profile is unreliable because of strong contamination \citep[][]{DeCia2016}. Thus, we excluded it from the calculation of the total N(PII). The new total column density is 12.88$\pm$0.28, which is lower than the previous value by 0.35 dex.

Among six DLAs for which there are measurements of P, four do not show significant enhancement ([P/Fe]$_{\rm nucl}\leq$0.30 dex), while the rest have values higher than 0.45 dex.  However, in all these cases the PII column densities were measured only from one line ($\lambda$ 1152$\AA$ or 963$\AA$), which is located in the Ly-$\alpha$ forest \citep[][]{DeCia2016}.  In general, it is necessary to fit multiple lines of the same ion simultaneously to obtain reliable measurements of the column density and avoid contaminations from other sources. In particular, contamination from HI lines are very likely within the Ly-$\alpha$ forest (bluer than Ly-$\alpha$). In addition, the systems with high \dvn\, are even more likely to be contaminated, because they have a broader velocity profile. Thus, we did not find the current data to be reliable enough to robustly assess the [P/Fe]$_{\rm nucl}$. Nevertheless, the methodology in this paper shows the potential for discovery in this field.

\subsection{Zinc}

Zn and Fe are often assumed to follow each other in nucleosytnhesis in studies of DLAs with metallicities between 1/100th of Solar and Solar, so that the observed gas-phase [Zn/Fe] is interpreted as a pure tracer of dust \citep[e.g.][]{DeCia2016, DeCia2018b, Konstantopoulou2022}. However, according to \citep[][]{Bensby2003, Nissen2011, Mishenina2011, Barbuy2015, Duffau2017}, the relation [Zn/Fe] versus [M/H] in stars in this metallicity range tends to behave partially like an [$\alpha$/Fe], but with a smaller amplitude. \citet{Sitnova2022} showed from non-LTE calculations for the MW stars that Zn and Mg do not strictly follow each other. [Zn/Fe] is decreasing with metallicity because of various contributions from core-collapse SNe and SNe Ia, and their different timescales (as for $\alpha$-elements). 

\begin{figure}
   \includegraphics[width=1.0\linewidth]{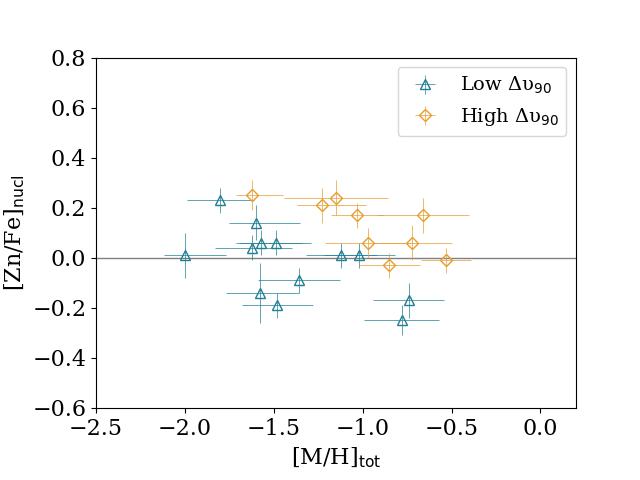}
  \caption{Zinc abundances in the ISM of DLAs for the golden sample for low-$\Delta v_{90}$ (blue open triangles) and high $\Delta v_{90}$ (orange open diamonds). }
\label{fig:Zn-DLAs}
\end{figure}

Our measurements of [Zn/Fe]\nucl\, are reliable only for the golden sample because Zn is not included in the fit for to the abundance patterns. Figure~\ref{fig:Zn-DLAs} shows the [Zn/Fe]\nucl\, with metallicity for the golden sample, where the strength of depletion is calculated only from $\alpha$-elements, without involving Zn. The distribution partially resembles the behaviour of [$\alpha$/Fe]\nucl, but with smaller amplitudes ([Zn/Fe]\nucl\, up to 0.2 dex). We observe a potential decrease of [Zn/Fe]\nucl, which is not unlike that of \citet{Sitnova2022}. The low- and high-$\Delta v_{90}$ subsamples form separate sequences:\ one at lower and one at higher metallicities. For the non-golden sample, it is not meaningful to analyse the [Zn/Fe]\nucl because it is distributed around zero by construction (see Fig. \ref{fig:Zn-DLAs-non-golden}).

\subsection{Additional caveats}
\label{sec:add_caveats}

This work is based on the assumption that we know well the refractory behaviour of different metals, namely, according to the refractory index $B2_X$. These coefficients were calculated by \citet{Konstantopoulou2022} and \citet{DeCia2016} based on the observed depletion sequences, that is, how the depletion vary with the overall amount of dust. This was derived after taking an assumption on the $\alpha$-element enhancement and Mn underabundance to focus on and characterise the properties of dust depletion. Our results on the $\alpha$-elements and Mn are independent from the initial assumptions of \citet{Konstantopoulou2022} and \citet{DeCia2016}, for the following reasons. First, this paper uses additional data with respect to \citet{Konstantopoulou2022} and \citet{DeCia2016}. Second, the enhancement of $\alpha$-elements and deficiency of Mn are observed for DLA systems with [Zn/Fe]$\sim0$ \citep[e.g. Fig. 3 of][]{DeCia2016}, so there is no assumption in this regime (the $\alpha$-plateau at low metallicities). Third, the assumed $\alpha$-knee results of \citet{Konstantopoulou2022} and \citet{DeCia2016} were at the same metallicity as for the MW ([M/H] $\sim-1.2$) and the same for all systems. These works made an overall statistical correction for $\alpha$-elements to the whole sample, without studying the details of individual systems. Eventually, the $\alpha$-knee that we find are at $[M/\mbox{H}]_{\rm tot}\sim-1.0$ and $[M/\mbox{H}]_{\rm tot}\sim-1.8$, different than previous assumptions. Finally, we measured different $\alpha$-knees for different samples of galaxies, which make physical sense. \\

\subsection{Testing the robustness of the results on the $\alpha$-elements.}
\label{sec:robustness}


To derive the depletion coefficients $B2_X$ for $\alpha$-elements, \citet{Konstantopoulou2022} applied corrections to account for the effect of $\alpha$-element enhancement. To do so, they adopted the shape of a standard nucleosynthetic curve taken from \citet{DeCia2016} (see also Section \ref{sec:alpha-elements}). To demonstrate that the assumption made by \citet{DeCia2016} and \citet{Konstantopoulou2022} has no (or minimal) impact on our present results, we tested our calculations using an alternative set of $B2_X$ coefficients; for instance, considering that  no assumption on $\alpha$-element nucleosynthesis had ever been made. For this purpose, we adopted the $B2_X$ coefficients of \citet{Konstantopoulou2022}, but before the correction for $\alpha$-element enhancement was made. The test coefficients should not be considered as more solid, because they describe dust depletion without taking any potential $\alpha$-element enhancements into account. Nevertheless, they are useful to check the robustness of our results and test that the assumptions that went into the calculation of the $B2_X$ does not substantially affect our results. 

With the test calculations, the values of $B2_X$ turn out to be systematically shifted downward by 0.11--0.20 depending on the element (S, Mg, O, Si, and Ti). With the test coefficients the $\alpha$-element enhancement is either the same or slightly higher compared to the reference values. However, the difference can become more prominent, up to 0.10 dex, if the slope of the dust depletion [Zn/Fe]$_{\rm fit}$ is steep. In average the difference of [$\alpha$/Fe]${\rm nucl}$ calculated with the test and standard $B2_X$ coefficients is 0.03 for both low- and high-$\Delta v_{90}$ subsamples. Despite these differences, we reproduced the same $\alpha$-element behaviour as in Fig. \ref{fig:alpha-met-DLAs}, but with slightly different fitting parameters: in case of test calculations, the position of the high-$\alpha$ knee is at [M/H]$_{\rm tot}$ lower by 0.02 and 0.09 dex for low- and high-\dvn\, subsamples, respectively.  The $\alpha$-plateau becomes slightly higher compared to the reference values by 0.1 and 0.05 dex for low- and high-$\Delta v_{90}$ subsamples, respectively. The two subsamples are still clearly separated and the assumption used by \citet{Konstantopoulou2022} has hardly any impact on our results.

For the golden sample, measurements of Ti are derived from the weak TiII $\lambda$1910 absorption line. If there is a contamination of the line, the column density of Ti could be overestimated and this may lead to an underestimation of the metallicity. We test our main results by deriving [M/H]\tot\, and [Zn/Fe]\fit\, for the golden sample in the same way as for the non-gold sample, namely, based on only iron-group elements (Fe, Ni, Cr) and Zn and without using Ti and other $\alpha-$elements. There are some differences in metallicities (0.15 dex in average) only for the high-$\Delta v_{90}$ subsample, while for the low-$\Delta v_{90}$ subsample the estimates are the same. Thus, our conclusions are holding regardless of the Ti measurements.

\section{Conclusions}

In this paper, we study abundance patterns of the neutral ISM in 110 QSO-DLA systems. We characterise the strength of depletion [Zn/Fe]${\rm _{fit}}$ and measure the total (dust + gas) metallicities [M/H]${\rm _{tot}}$. In addition, from the deviations of the observed abundances from the linear fits to the abundance patterns, we derive [X/Fe]${\rm _{nucl}}$ for O, Mg, Si, S, Ti, Cr, Fe, Ni, Zn, P, and Mn. These are relative abundances with respect to Fe, after taking dust depletion into account and, thus, directly comparable to stellar relative abundances. 
We analyse the behaviour of [$\alpha$/Fe]${\rm _{nucl}}$, [Mn/Fe]${\rm _{nucl}}$, and [Zn/Fe]${\rm _{nucl}}$ depending on the total metallicity.

The main analysis is based on 24 QSO-DLAs (the golden sample) for which there are measurements of column densities of Ti and at least one other $\alpha$-element. We separated our golden sample into two groups, one with $\Delta v_{90}$ < 100 \kms\, (low $\Delta v_{90}$) and the other with $\Delta v_{90}$ > 100 \kms\, (high $\Delta v_{90}$). This separation is aimed at creating two subsamples of galaxies: one with a higher and one with lower average stellar mass. In addition, we made a minimal analysis for a non-golden sample for which the depletion by dust and the total metallicity have been obtained from a linear fitting abundances of Zn and the iron-group elements Fe, Cr, and Ni.

For the golden sample, we found that less massive galaxies show an $\alpha$-element knee at lower metallicities than more massive galaxies. If this collective behaviour can be interpreted as for individual systems, this would suggest that more massive and metal-rich systems evolve to higher metallicities before the contribution of SN-Ia levels out the [$\alpha$/Fe] enhancement created by core-collapse SNe. This is possibly explained by different SFR in galaxies of different masses.

For the golden sample, there is a clear manganese under-abundance at about constant level up to [M/H]${\rm _{tot}}$ = $-$0.53 dex with average values of [Mn/Fe]${\rm _{nucl}}$ equal to $-$0.33$\pm$0.07 and $-$0.39$\pm$0.07 for the high- and low-$\Delta v_{90}$ subsamples, respectively. This is not fully consistent with our expectations since (according to the behaviour of $\alpha$-elements) in some QSO-DLAs, SNe Ia contribute to abundance pattern that, in turn, should lead to increase of [Mn/Fe]${\rm _{nucl}}$. It is only for the non-golden sample that [Mn/Fe]${\rm _{nucl}}$ increases starting from [M/H]${\rm _{tot}}$ = $-$0.69 dex.

We traced slight effects in the behaviour of [Zn/Fe]$_{\rm nucl}$ with the total metallicity that resemble the behaviour of [$\alpha$/Fe]$_{\rm nucl}$ but with a smaller amplitude. This is in agreement with several works on stellar relative abundances \citep[][]{Bensby2003, Nissen2011, Mishenina2011, Barbuy2015, Duffau2017}. A systematic bias by [M/H]${\rm _{tot}}$ between high- and low-$\Delta v_{90}$ subsamples is observed. These effects can be obtained only for the golden sample. Otherwise, by construction of the method [Zn/Fe]$_{\rm nucl}$ should be approximately Solar.

Measurements of phosphorus should be treated with caution, since in all cases the column densities were measured only from one line ($\lambda$ 1152$\AA$ or 963$\AA$), which is located in the Ly-$\alpha$ forest (bluer than Ly-$\alpha$). Probably, because of this, all DLAs in the high-$\Delta v_{90}$ subsample have inexplicably high values of [P/Fe]$_{\rm nucl}$, higher than 0.45 dex. A wider velocity profile is more likely to be contaminated and lead to erroneous measurement. The low-$\Delta v_{90}$ DLAs do not show significant enhancement of P, with [P/Fe]$_{\rm nucl}\leq$0.15 dex.

Studying the chemical properties of the neutral ISM in a sample of QSO-DLAs enables us to trace the collective behaviour of element abundances that can be interpreted as for individual systems. A homogeneous set of $\alpha$-element abundances (Ti and at least one other $\alpha$-element) provides a powerful tool for estimating the strength of depletion [Zn/Fe]\fit\, and the total metallicity [M/H]\tot\, in the ISM of QSO-DLAs, as well as the more subtle deviations in the relative abundances due to the nucleosynthesis of specific stellar populations. This opens a new window into the study of the chemical evolution of distant galaxies.

\begin{acknowledgements} 
We thank the anonymous referee for the useful and constructive comments that improved this manuscript. A.V.,  A.D.C., C.K., J.K.K. and T.R.H. acknowledge support by the Swiss National Science Foundation under grant 185692 funding the ``Interstellar One'' project. 
\end{acknowledgements}

\bibliographystyle{aa} 

\begin{appendix}

\section{Element abundances in the golden sample}
\label{app:abund}

\renewcommand{\arraystretch}{1.2}
\setlength{\tabcolsep}{2pt}
\begin{table*}[]
    \small
    \centering
    \caption{Element abundances derived after taking into account the dust depletion.}
    \begin{tabular}{lcccccccccccc}
\hline
QSO & [O/Fe]${\rm _{nucl}}$ & [Mg/Fe]${\rm _{nucl}}$ & [Si/Fe]${\rm _{nucl}}$ & [S/Fe]${\rm _{nucl}}$  & [Ti/Fe]${\rm _{nucl}}$  & [Cr/Fe]${\rm _{nucl}}$  & [Ni/Fe]${\rm _{nucl}}$  & [P/Fe]${\rm _{nucl}}$  & [Mn/Fe]${\rm _{nucl}}$  & [Zn/Fe]${\rm _{nucl}}$ \\
\hline
Q0058$-$292  &...& 0.26$\pm$0.04 & 0.23$\pm$0.05 & 0.27$\pm$0.04 & 0.29$\pm$0.12 & 0.07$\pm$0.05 & 0.00$\pm$0.09  &... &... &$-$0.09$\pm$0.05  \\ 
Q0100$+$130  &...& 0.31$\pm$0.07  &...& 0.32$\pm$0.04 & 0.32$\pm$0.12 & 0.01$\pm$0.04  &...& $-$0.05$\pm$0.06 & $-$0.48$\pm$0.07  & 0.04$\pm$0.05  \\ 
Q0102$-$190a  &... &... &...& 0.17$\pm$0.04 & 0.37$\pm$0.13 & 0.06$\pm$0.05 & $-$0.10$\pm$0.08  &...& $-$0.49$\pm$0.07  & 0.01$\pm$0.09  \\ 
Q0405$-$443a  &... &...& 0.04$\pm$0.04  &...& 0.04$\pm$0.06 & 0.00$\pm$0.05 & $-$0.10$\pm$0.05  &...& $-$0.41$\pm$0.07  & $-$0.25$\pm$0.06  \\ 
Q0405$-$443c  &... &...& 0.19$\pm$0.04 & 0.19$\pm$0.04 & 0.18$\pm$0.08 & 0.03$\pm$0.05 & $-$0.09$\pm$0.05 & $-$0.02$\pm$0.06 & $-$0.38$\pm$0.07  & 0.01$\pm$0.05  \\ 
Q0528$-$250a  &... &...& 0.41$\pm$0.04 & 0.33$\pm$0.04 & 0.45$\pm$0.09 & 0.03$\pm$0.05 & 0.05$\pm$0.05  &...& $-$0.30$\pm$0.08  & 0.25$\pm$0.06  \\ 
Q0528$-$250b  &...& 0.05$\pm$0.04 & 0.33$\pm$0.04 & 0.10$\pm$0.04 & 0.05$\pm$0.08 & $-$0.07$\pm$0.04 & 0.01$\pm$0.05  &...& $-$0.31$\pm$0.06  & $-$0.03$\pm$0.05  \\ 
Q0841$+$129a  &... &... &...& 0.13$\pm$0.06 & 0.13$\pm$0.08  &...& $-$0.10$\pm$0.05  &...& $-$0.43$\pm$0.07  &$-$0.14$\pm$0.12  \\ 
Q1117$-$134  &... &...& 0.20$\pm$0.04  &...& 0.29$\pm$0.08 & $-$0.03$\pm$0.06  & <$-$0.10 &... &... & 0.14$\pm$0.07  \\ 
Q1157$+$014  &...& 0.25$\pm$0.05 & 0.22$\pm$0.04  &...& 0.23$\pm$0.06 & 0.02$\pm$0.04 & $-$0.09$\pm$0.05 & 0.15$\pm$0.07 & $-$0.36$\pm$0.06  & 0.06$\pm$0.05  \\ 
Q1223$+$178  &...& 0.41$\pm$0.06 & 0.12$\pm$0.04 & 0.17$\pm$0.04 & 0.20$\pm$0.08 & $-$0.03$\pm$0.04 & $-$0.16$\pm$0.07  &...& $-$0.48$\pm$0.07  & $-$0.19$\pm$0.05  \\ 
Q2206$-$199a  &...& 0.14$\pm$0.06 & 0.25$\pm$0.04 & 0.23$\pm$0.04 & 0.22$\pm$0.06 & $-$0.04$\pm$0.04 & 0.01$\pm$0.05 & 0.46$\pm$0.07 & $-$0.38$\pm$0.07  & $-$0.01$\pm$0.05  \\ 
Q2243$-$605  &... &...& 0.31$\pm$0.04 & 0.34$\pm$0.04 & 0.34$\pm$0.07 & $-$0.06$\pm$0.05 & 0.06$\pm$0.05 & 0.54$\pm$0.06 & $-$0.26$\pm$0.07  & 0.17$\pm$0.05  \\ 
SDSS1249$-$0233  &... &... &...& 0.27$\pm$0.06 & 0.27$\pm$0.12 & 0.13$\pm$0.05 & 0.01$\pm$0.07  &...& $-$0.40$\pm$0.07  & 0.24$\pm$0.07  \\ 
Q2230$+$025  &... &...& 0.22$\pm$0.06 & 0.22$\pm$0.06 & 0.22$\pm$0.11 & $-$0.10$\pm$0.06 & 0.13$\pm$0.07  &...& $-$0.31$\pm$0.07  & 0.06$\pm$0.07  \\ 
HE1104$-$1805 & 0.32$\pm$0.20  &...& 0.30$\pm$0.04  &...& 0.30$\pm$0.09 & $-$0.00$\pm$0.04 & $-$0.15$\pm$0.05  &...& $-$0.18$\pm$0.15  & 0.01$\pm$0.05  \\ 
Q0027$-$1836  &...& 0.73$\pm$0.06 & 0.44$\pm$0.05 & 0.35$\pm$0.04 & 0.39$\pm$0.08 & 0.09$\pm$0.04 & $-$0.07$\pm$0.05  &...& $-$0.29$\pm$0.07  & 0.23$\pm$0.05  \\ 
Q1210$+$17  &... &...& 0.08$\pm$0.05 & 0.07$\pm$0.04 & 0.06$\pm$0.10 & $-$0.04$\pm$0.05 & $-$0.14$\pm$0.07  &...& $-$0.44$\pm$0.06  & $-$0.17$\pm$0.07  \\ 
FJ0812$+$32  &... &...& 0.20$\pm$0.04  &...& 0.25$\pm$0.07 & 0.03$\pm$0.04 & $-$0.09$\pm$0.05  &...& $-$0.32$\pm$0.07  & 0.06$\pm$0.05  \\ 
J1135$-$0010  &... &...& 0.36$\pm$0.05  &...& 0.36$\pm$0.07 & 0.08$\pm$0.04 & 0.12$\pm$0.05  &...& $-$0.35$\pm$0.06  & 0.06$\pm$0.06  \\ 
Q2231$-$00  &... &...& 0.37$\pm$0.05 & 0.62$\pm$0.15 & 0.44$\pm$0.10 & $-$0.15$\pm$0.06 & $-$0.11$\pm$0.07  &...& $-$0.32$\pm$0.07  & 0.21$\pm$0.07  \\ 
J1200$+$4015  &... &... &...& 0.31$\pm$0.06 & 0.31$\pm$0.07 & $-$0.10$\pm$0.06 & 0.06$\pm$0.07  &... &... & 0.17$\pm$0.07  \\ 
Q1210$+$175  &...& 0.28$\pm$0.09  &...& 0.02$\pm$0.08 & 0.11$\pm$0.06  &...& $-$0.02$\pm$0.05  &...& $-$0.33$\pm$0.08  &... \\ 
Q1232$+$082  &...& 0.60$\pm$0.09 & 0.53$\pm$0.08 & 0.39$\pm$0.08 & 0.46$\pm$0.10  &...& $-$0.03$\pm$0.06 & 0.26$\pm$0.18 & $-$0.14$\pm$0.13  &... \\ 
\hline
\end{tabular}
\label{tab:abund_golden}
\end{table*}

\begin{figure*}[h!]
 \includegraphics[width=1.0\textwidth]{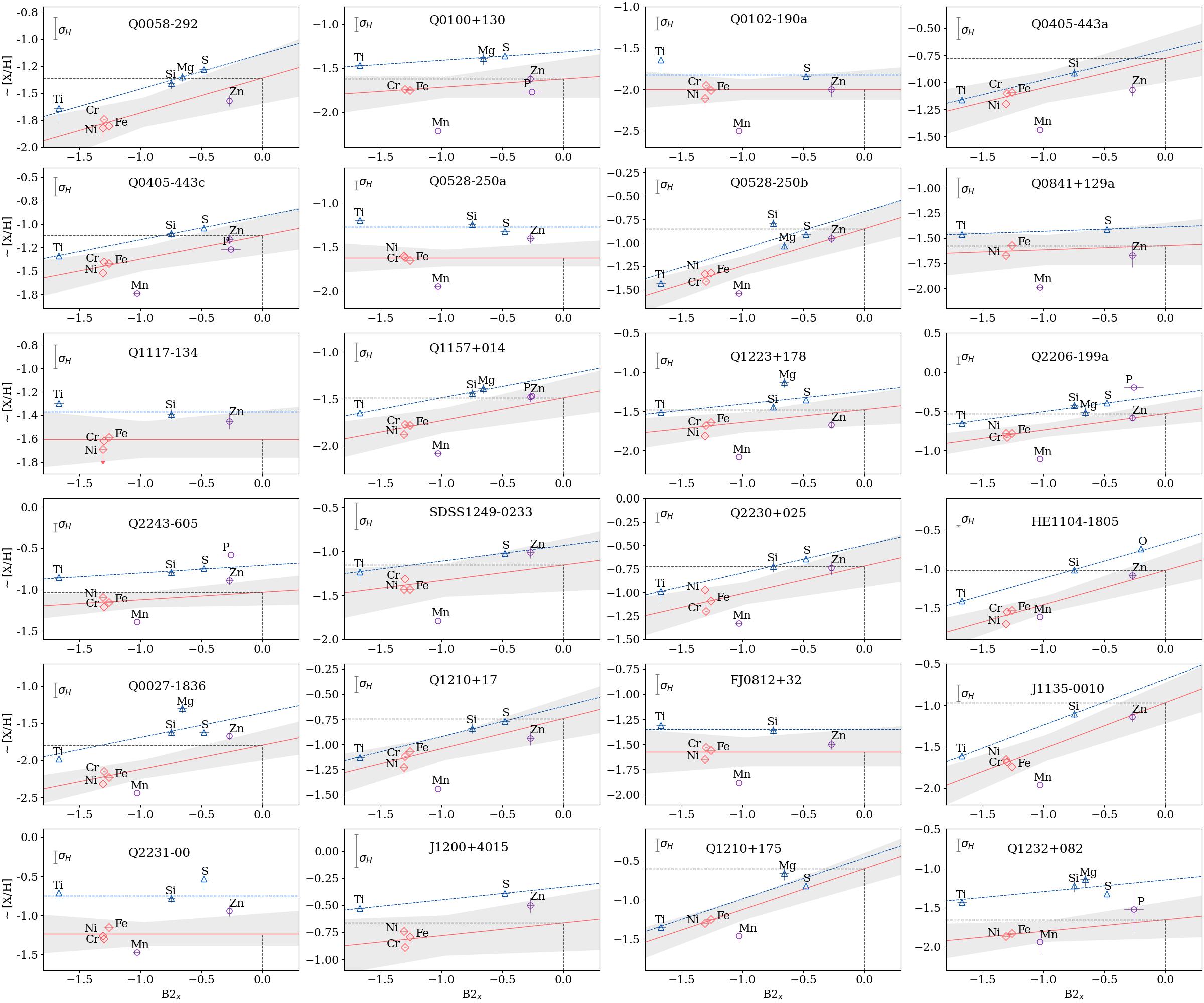} 
\caption{Abundance patterns for the golden sample of QSO-DLA (see Sec. \protect\ref{sec:data}). Blue dashed curve is fitting to the $\alpha$-elements, shown by blue open triangles. Red curve is shifted blue one to the value averaged through Ni, Cr, and Fe. The intersection of the gray dashed lines at $B2_X = 0$ shows the metallicity [M/H]$_{\rm tot}$ of each system. Gray areas show 1$\sigma$ confidence intervals.}
\label{fig:XH_B2X_golden1}
\end{figure*}

\section{Element abundances in the non-golden sample}
\label{app:abund_non-golden}

\begin{figure*}
  \centering 
  \subfloat{ \includegraphics[width=1.0\textwidth]{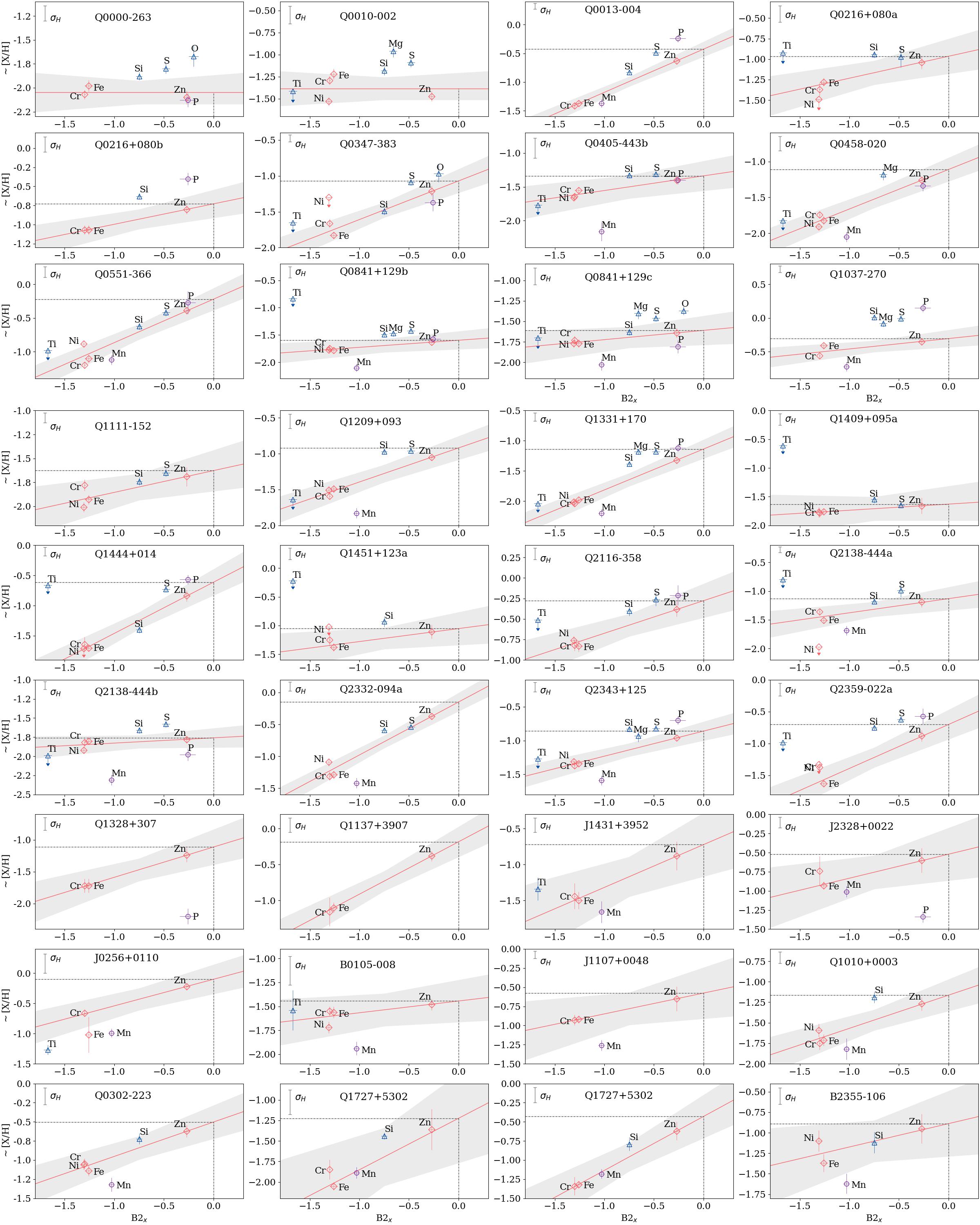} }%
  \caption{Abundance patterns for the non-golden sample of QSO-DLA (see Sec. \protect\ref{sec:data}). Total metallicity [M/H]\tot\, and the dust depletion [Zn/Fe]\fit\, have been derived from fitting iron-group elements and Zn.}
  \label{fig:XH_B2X_non-golden1}
\end{figure*}

\begin{figure*}
  \ContinuedFloat 
  \centering 
  \subfloat{\includegraphics[width=1.0\textwidth]{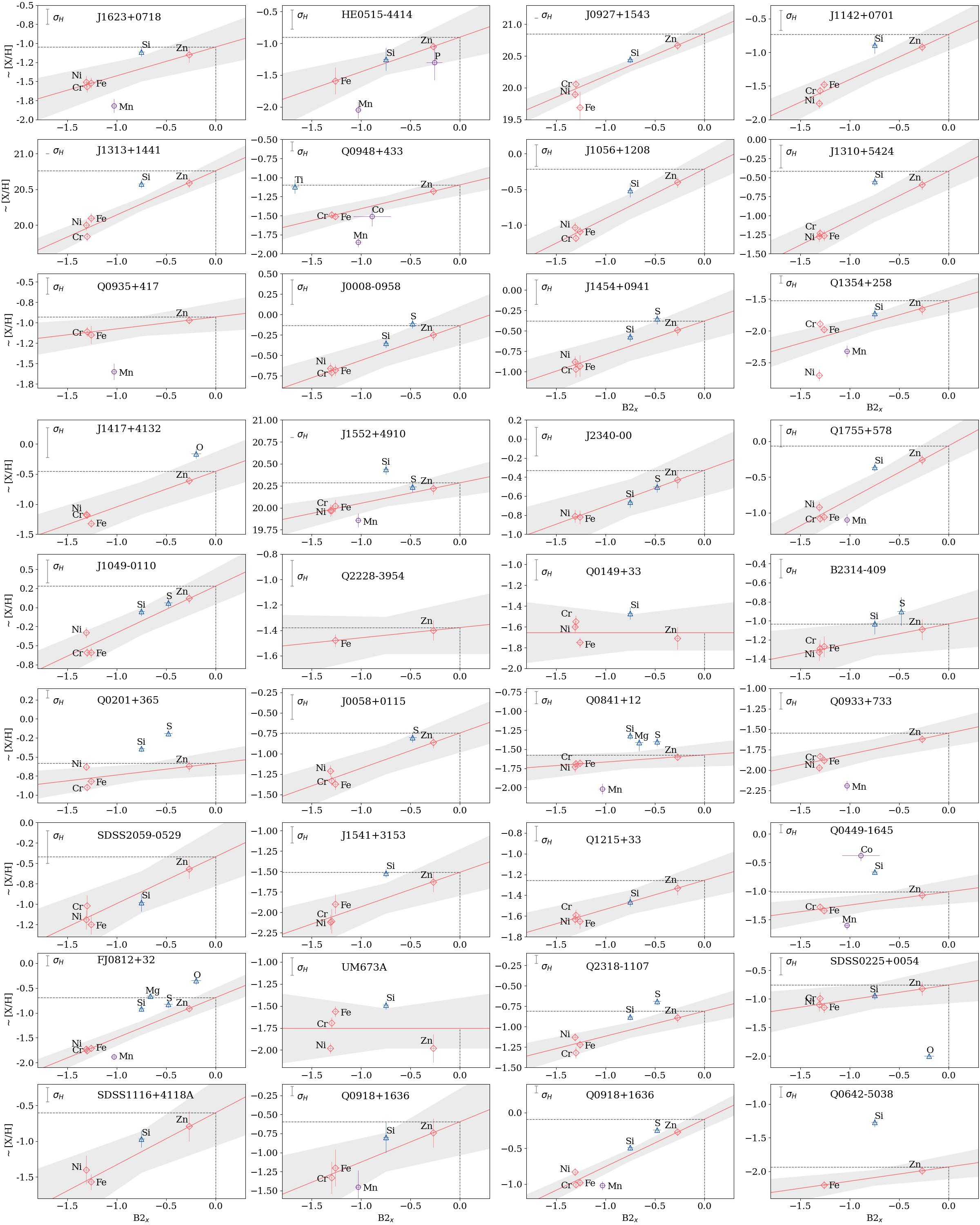}}%
  \caption{continued.}
  \label{fig:XH_B2X_non-golden2}
\end{figure*} 

\begin{figure*}
  \ContinuedFloat 
  \centering 
  \subfloat{ \includegraphics[width=1.0\textwidth]{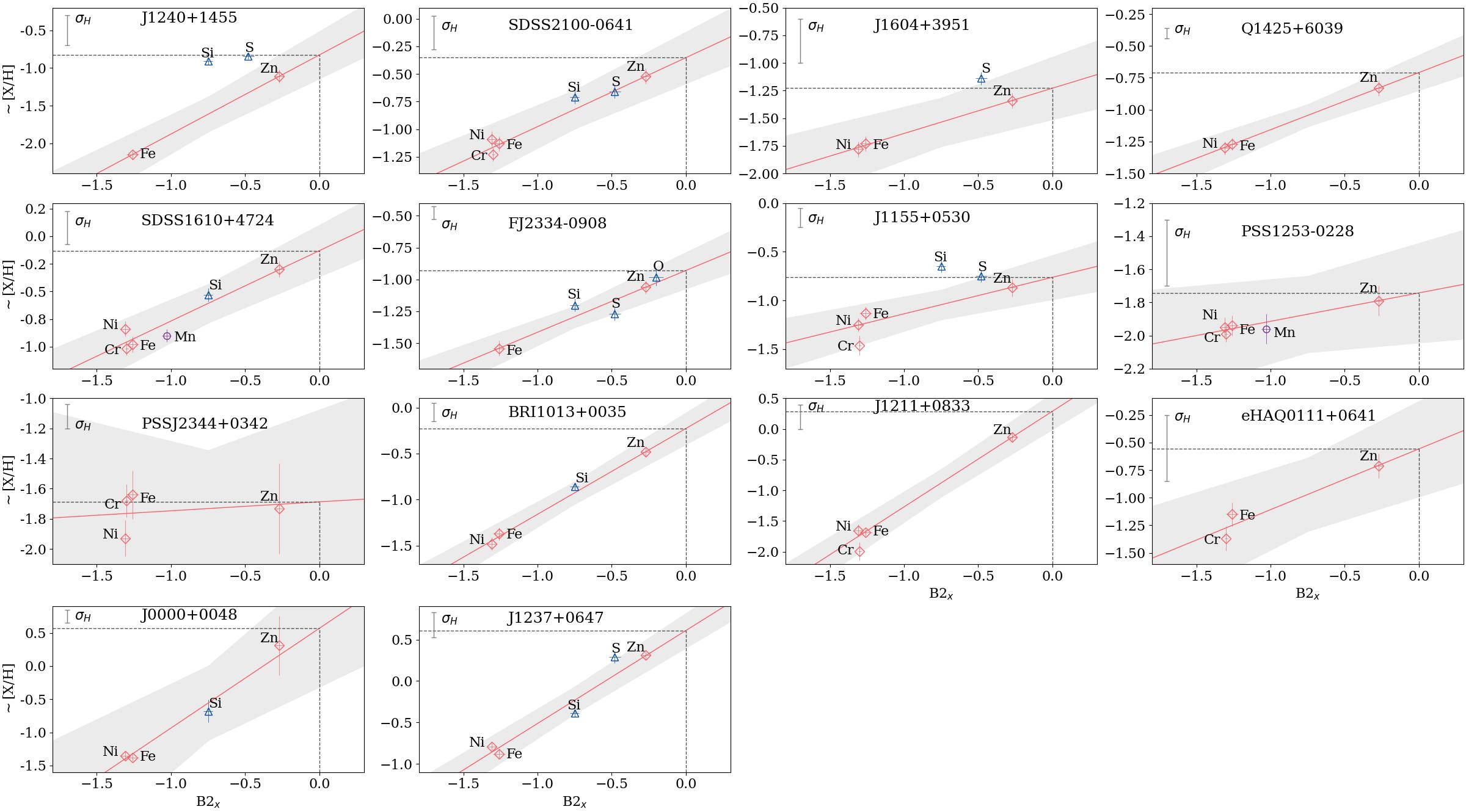}}%
  \caption{continued.}
  \label{fig:XH_B2X_non-golden3}
\end{figure*} 

\section{Dwarf galaxies}
\label{app:dwarf_gal_fit}

\begin{figure*}[h!]
 \includegraphics[width=1.0\textwidth]{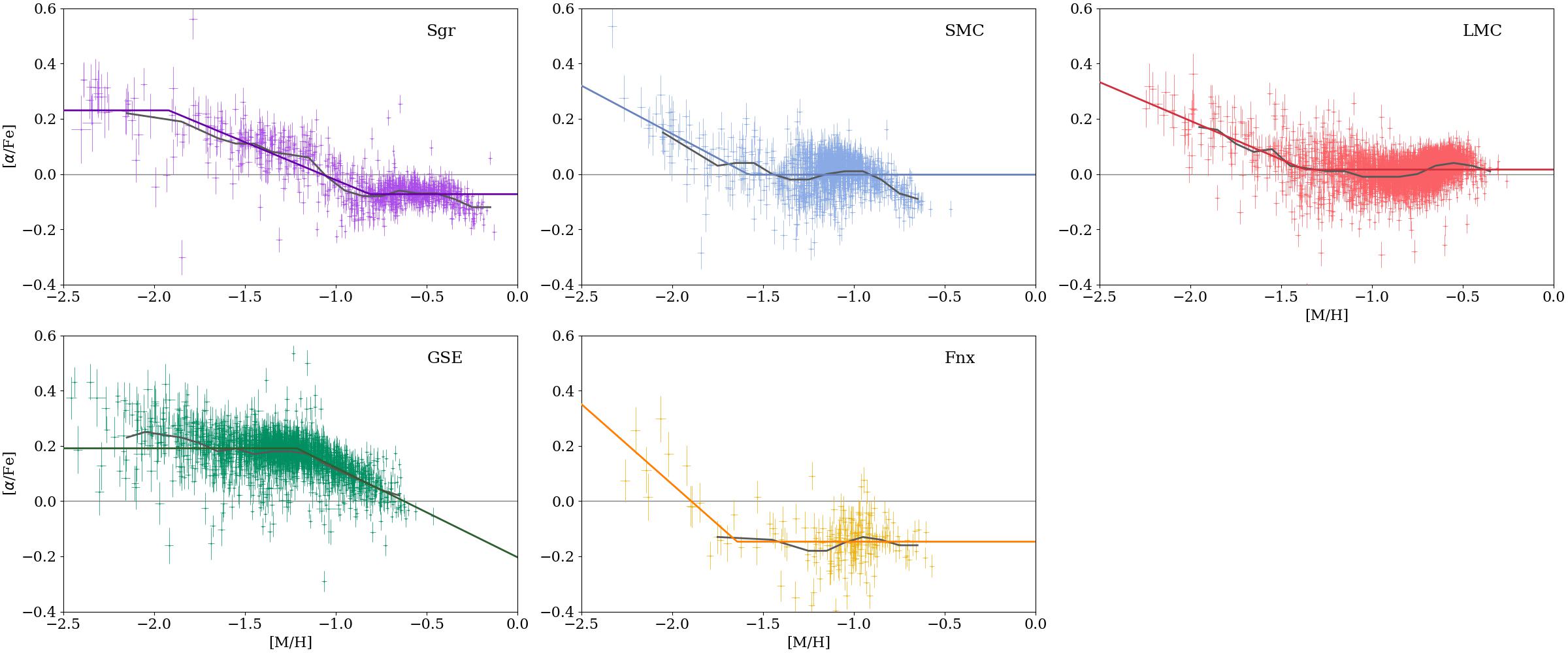} 
\caption{ Chemical abundances of $\alpha$-elements taken from the APOGEE DR17 catalog for dwarf galaxies (color points with $x, y$ error bars), with \cite{Hasselquist2021} selection. Color solid curves show the approximation of the data points by piecewise functions consisting of two or three straight lines representing the average behaviour of $\alpha$-elements relative to iron. Dark gray solid curves have been obtained by averaging the data in each metallicity bin. Typically, the bin width is equal to 0.1 dex, but in some ranges where there are too few stars, it is increased up to 0.2 dex. For comparison, the behaviour typical for the MW is shown by gray dashed curve (\cite{McWilliam1997}).}
\label{fig:alpha-galaxies}
\end{figure*}

\section{Fitting the data}

\begin{figure*}
\centering
   \includegraphics[width=0.33\linewidth]{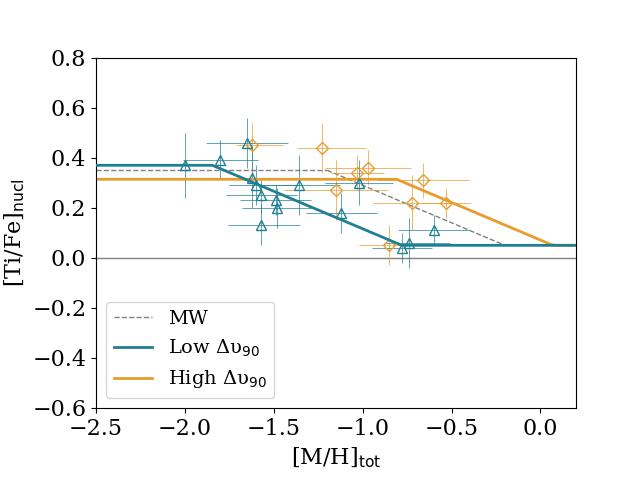}
   \includegraphics[width=0.33\linewidth]{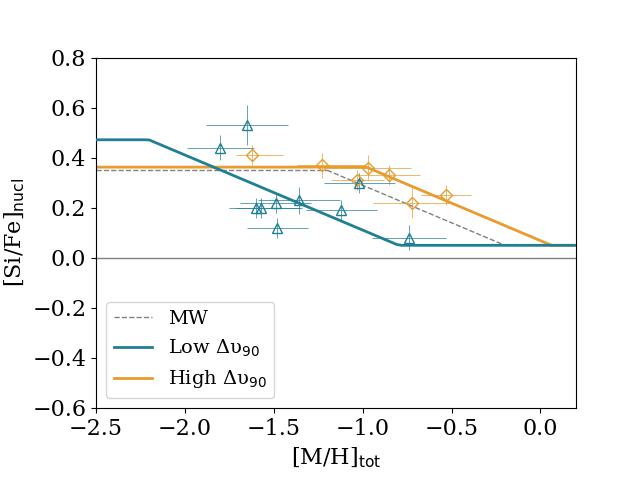}
   \includegraphics[width=0.33\linewidth]{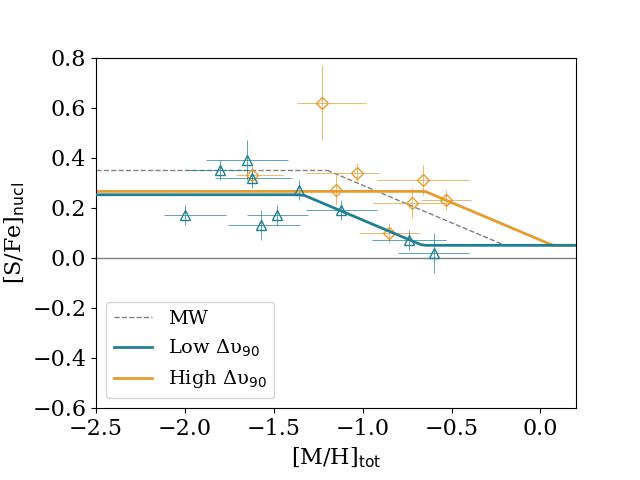}
  \caption{ [X/Fe]\nucl\, for three $\alpha$-elements Ti, Si and S. The data are fitted by three-piecewise functions as for Case 2 to find the level of the high-$\alpha$ plateau and the position of the high-$\alpha$ knee.}
\label{fig:S-Si-Ti-met-DLAs-add}
\end{figure*}

\section{Abundances of zinc for the non-golden sample}

\begin{figure*}
\centering
  \includegraphics[width=0.5\linewidth]{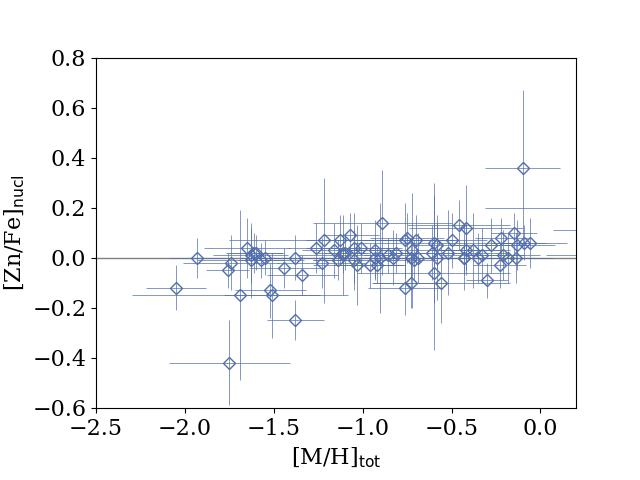}
  \caption{Zinc abundances in the ISM of QSO-DLAs for the non-golden sample.}
\label{fig:Zn-DLAs-non-golden}
\end{figure*}

\section{New column densities}

\begin{figure}
\centering
  \includegraphics[width=1.0\linewidth]{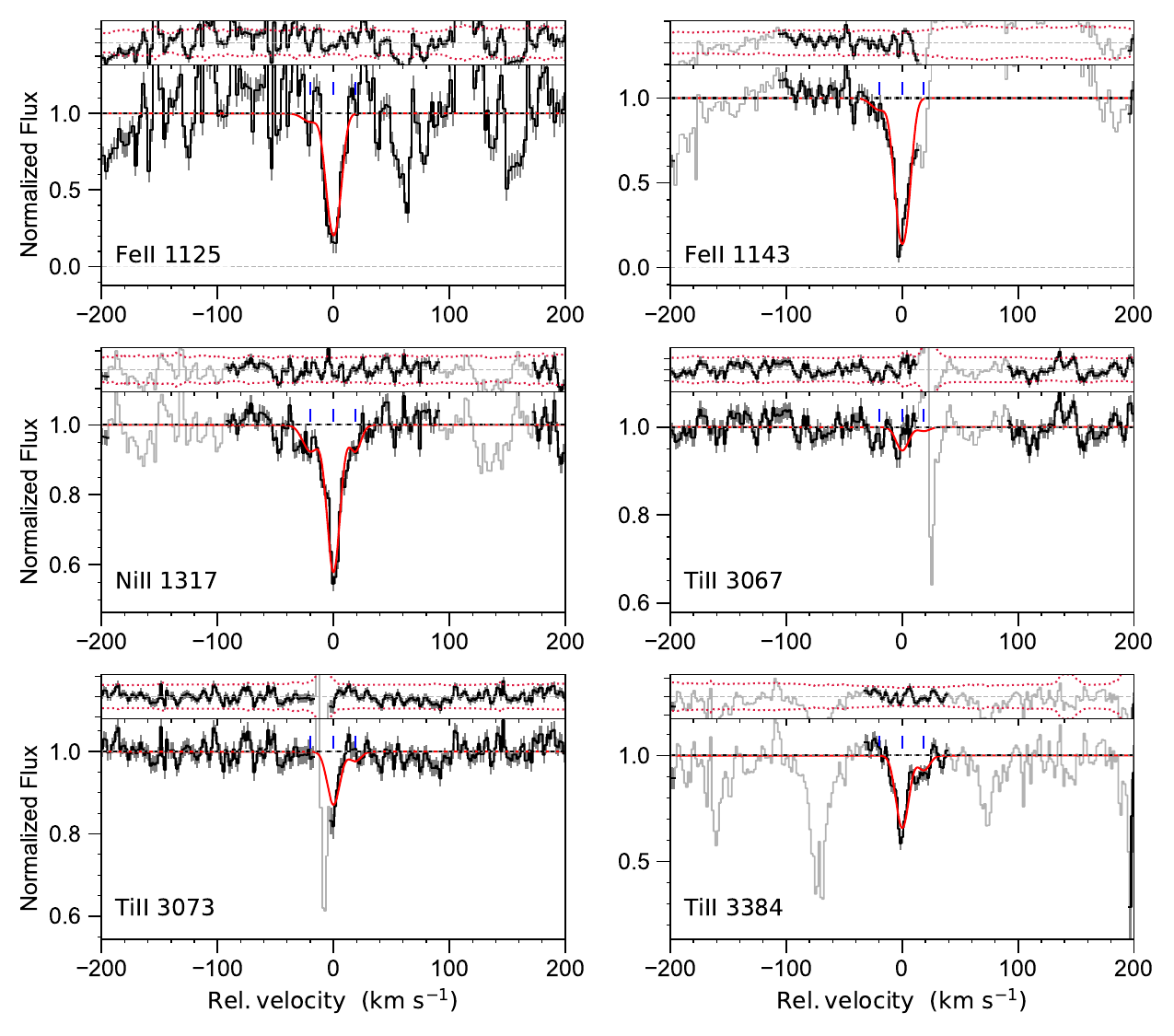}
  \caption{Velocity profiles of selected low-ionisation transition lines from the DLA system at $z_{\rm abs}$ = 1.892 towards Q 1210+175.}
\label{fig:col_dens_Q1210}
\end{figure}

\begin{figure}
\centering
  \includegraphics[width=1.0\linewidth]{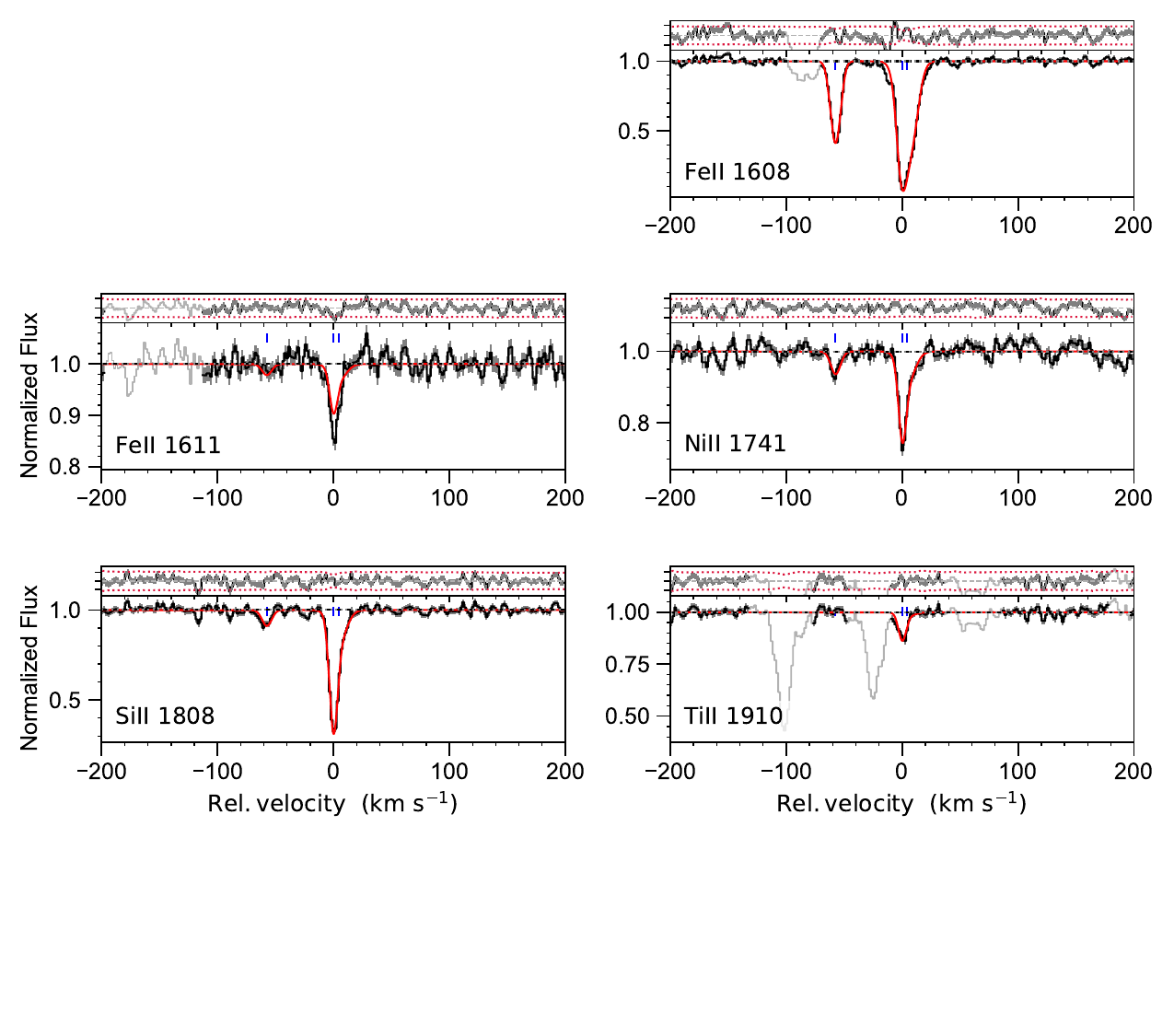}
  \caption{Velocity profiles of selected low-ionisation transition lines from the DLA system at $z_{\rm abs}$ = 2.338 towards Q 1232+082.}
\label{fig:col_dens_Q1232}
\end{figure}

\renewcommand{\arraystretch}{1.0}
\begin{table*}[]
    \centering
    \begin{tabular}{lcccc}
\hline
QSO & $z_{\rm abs}$ & log\,$N$(FeII) & log\,$N$(NiII) & log\,$N$(TiII) \\
\hline
Q0102$-$190b & 2.92648 & 13.80$\pm$0.02 & $<$13.05 & $<$12.38 \\ 
Q0112+030 & 2.42299 & 14.85$\pm$0.02 & 13.57$\pm$0.06 & 12.48$\pm$0.16 \\ 
Q0112$-$306a & 2.41844 & 13.42$\pm$0.04 & ... & $<$12.28 \\ 
Q0112$-$306b & 2.70163 & 14.81$\pm$0.03 & 13.65$\pm$0.01 & ... \\ 
Q0135$-$273a & 2.10735 & 14.58$\pm$0.03 & $<$13.35 & $<$12.50 \\ 
Q0135$-$273b & 2.80004 & 14.77$\pm$0.03 & 13.46$\pm$0.06 & $<$12.51 \\ 
Q0336$-$017 & 3.06209 & 14.79$\pm$0.02 & 13.42$\pm$0.05 & $<$11.92 \\ 
Q0450$-$131 & 2.06658 & 14.24$\pm$0.02 & 13.34$\pm$0.04 & $<$12.65 \\ 
Q0913+072 & 2.61843 & 13.10$\pm$0.02 & $<$12.42 & $<$12.02 \\ 
Q1036$-$229 & 2.77779 & 14.76$\pm$0.01 & 13.71$\pm$0.02 & $<$12.16 \\ 
Q1108$-$077b & 3.60767 & 13.88$\pm$0.02 & $<$12.35 & $<$12.25 \\ 
Q1210+175 & 1.89177 & 14.90$\pm$0.02 & 13.67$\pm$0.02 & 12.31$\pm$0.03 \\ 
Q1232+082 & 2.33771 & 14.52$\pm$0.01 & 13.30$\pm$0.03 & 12.43$\pm$0.09 \\ 
Q1337+113a & 2.50792 & 13.40$\pm$0.02 & $<$12.85 & $<$12.55 \\ 
Q1337+113b & 2.79584 & 14.33$\pm$0.02 & 13.20$\pm$0.06 & ... \\ 
Q1340$-$136 & 3.11835 & 13.93$\pm$0.02 & 12.89$\pm$0.05 & ... \\ 
Q1409+095b & 2.45593 & 13.74$\pm$0.02 & $<$12.66 & $<$12.32 \\ 
Q1451+123b & 2.46921 & 13.38$\pm$0.02 & $<$13.22 & $<$12.58 \\ 
Q1451+123c & 3.17112 & 13.35$\pm$0.09 & 13.30$\pm$0.06 & $<$13.11 \\ 
Q2059$-$360a & 2.50734 & 13.50$\pm$0.03 & $<$12.79 & $<$12.29 \\ 
Q2059$-$360b & 3.08261 & 14.48$\pm$0.02 & 13.02$\pm$0.06 & $<$12.29 \\ 
Q2152+137b & 3.31599 & 14.50$\pm$0.02 & $<$12.88 & $<$12.51 \\ 
Q2206$-$199b & 2.07622 & 13.34$\pm$0.01 & 12.24$\pm$0.09 & $<$11.59 \\ 
Q2230+025 & 1.86427 & 15.35$\pm$0.01 & 14.04$\pm$0.07 & $<$12.45 \\ 
Q2231$-$002 & 2.06615 & 14.92$\pm$0.01 & 13.61$\pm$0.01 & ... \\ 
Q2332$-$094b & 3.05722 & 14.37$\pm$0.01 & 13.19$\pm$0.11 & $<$12.51 \\ 
Q2344+125 & 2.53787 & 14.01$\pm$0.01 & 12.58$\pm$0.08 & $<$12.30 \\ 
Q2348$-$011a & 2.42695 & 14.83$\pm$0.01 & 13.74$\pm$0.02 & $<$12.52 \\ 
Q2348$-$011b & 2.61473 & 14.65$\pm$0.01 & 13.34$\pm$0.05 & ... \\ 
Q2359$-$022b & 2.15368 & 13.78$\pm$0.04 & $<$13.22 & $<$12.60 \\ 
    \end{tabular}
    \caption{Column densities for the DLAs that are absent in the list compiled by \citet{Konstantopoulou2022}.}
    \label{tab:col_dens}
\end{table*}

\end{appendix}
\end{document}